\documentclass[aps,prl,twocolumn,10pt,longbibliography]{revtex4-1}
\usepackage{amsmath}
\usepackage{amsfonts}
\usepackage{amssymb}
\usepackage{mathtools}
\usepackage[colorlinks=true,citecolor=blue,linkcolor=blue,urlcolor=blue]{hyperref}
\usepackage{bm}
\usepackage{times}
\usepackage{cases}
\usepackage{wasysym}
\usepackage[version=4]{mhchem}
\usepackage{graphicx}
\usepackage[caption=false]{subfig}
 
\begin{document}
\title{Is spontaneous vortex generation in superconducting 4Hb-TaS$_2$ from vison-vortex nucleation\\ 
with $\mathbb{Z}_2$ topological order?}
\author{Rui Leonard Luo}
\affiliation{Department of Physics and HKU-UCAS Joint Institute for Theoretical and Computational Physics at Hong Kong, 
The University of Hong Kong, Hong Kong, China}
\author{Gang V. Chen}
 \email{chenxray@pku.edu.cn}
\affiliation{International Center for Quantum Materials, School of Physics, Peking University, Beijing 100871, China}
\affiliation{Collaborative Innovation Center of Quantum Matter, Beijing 100871, China}

\date{\today}
    
\begin{abstract}
We propose the superconducting van der Waals material 4Hb-TaS$_2$ to realize the $\mathbb{Z}_2$ 
topological order and interpret the recent discovery of the spontaneous vortex generation in 4Hb-TaS$_2$
as the vison-vortex nucleation.  For the alternating stacking of metallic/superconducting and 
Mott insulating layers in 4Hb-TaS$_2$, we expect the local moments in the Mott insulating 1T-TaS$_2$ 
layer to form the $\mathbb{Z}_2$ topological order. The spontaneous vortex generation in 4Hb-TaS$_2$ 
is interpreted from the transition or nucleation between the superconducting vortex and 
the $\mathbb{Z}_2$ vison in different phase regimes. Differing from the single vison-vortex 
nucleation in the original Senthil-Fisher's cuprate proposal, we consider such nucleation process 
between the superconducting vortex lattice and the vison crystal. We further propose experiments 
to distinguish this proposal with the $\mathbb{Z}_2$ topological order from the chiral spin liquid scenarios. 
\end{abstract}

\maketitle

% detecting topological order and fractionalization 

\noindent{\emph{\bf{1. Introduction}}}\\
Detecting the intrinsic topological orders and the emergent electron fractionalization is an 
interesting and challenging task. There have been a tremendous effort to identify 
topological orders and the associated electron fractionalization~\cite{Wen_2013}. 
Among these efforts, symmetry has been quite useful in both classifying distinct 
topological orders and the identifications of measurable physical 
quantities~\cite{PhysRevB.65.165113,PhysRevB.87.155115,PhysRevB.87.104406,Kitaev_2006}. 
One important consequence of the symmetries in topological orders is  
the quantum number fractionalization, and one such application is the charge fractionalization 
in the fractional quantum Hall liquid~\cite{PhysRevLett.50.1395,PhysRevLett.53.722}. 
The crystallographic symmetry has also proved to be quite 
useful in identifying the crucial features in the spectrum of the 
anyonic quasiparticles~\cite{PhysRevB.65.165113,PhysRevB.87.104406}, 
and this is related to the crystal symmetry fractionalization in the topologically ordered phases. 
Another interesting effort is to trace the genesis of various exotic quasiparticles in the 
proximate phases~\cite{PhysRevB.62.7850,PhysRevLett.86.292,PhysRevB.61.6307,PhysRevB.60.1654,PhysRevB.71.125102}. 
This is of the fundamental importance to clarify the relationship between 
the topological phases and the proximate conventional states and
to understand the nature of the phase transitions~\cite{PhysRevB.71.125102,Senthil_2015,PhysRevB.94.205107}. 
More specifically,
in the context of the cuprates where the topological order was suggested and remains
elusive, Senthil and Fisher proposed a creative scheme to trace the nucleation 
between the vison in $\mathbb{Z}_2$ topological order and 
the superconducting vortex~\cite{PhysRevLett.86.292}. Although this vortex-memory effect 
was found to be absent in cuprates~\cite{Bonn2001},
this idea remains to be an ingenious attempt to detect the 
topological order beyond fractional quantum Hall liquids. 
 In this work,
we propose the superconducting van der Waals material 4Hb-TaS$_2$~\cite{Persky2022,Shen_2022}
to be a physical platform for the $\mathbb{Z}_2$ topological order,
and interpret the recently-discovered magnetic memory effect and 
the spontaneous vortex generation~\cite{Persky2022} 
as the vison-vortex nucleation due to the $\mathbb{Z}_2$ topological order. 
Our proposal for 4Hb-TaS$_2$ is a vortex lattice version of the Senthil-Fisher 
single-vortex proposal for cuprates, and bridges the puzzling experiments 
in 4Hb-TaS$_2$ with the forefront of topological orders and their detection. 
 \\

\begin{figure}[b] 
	\centering
	\includegraphics[width=7cm]{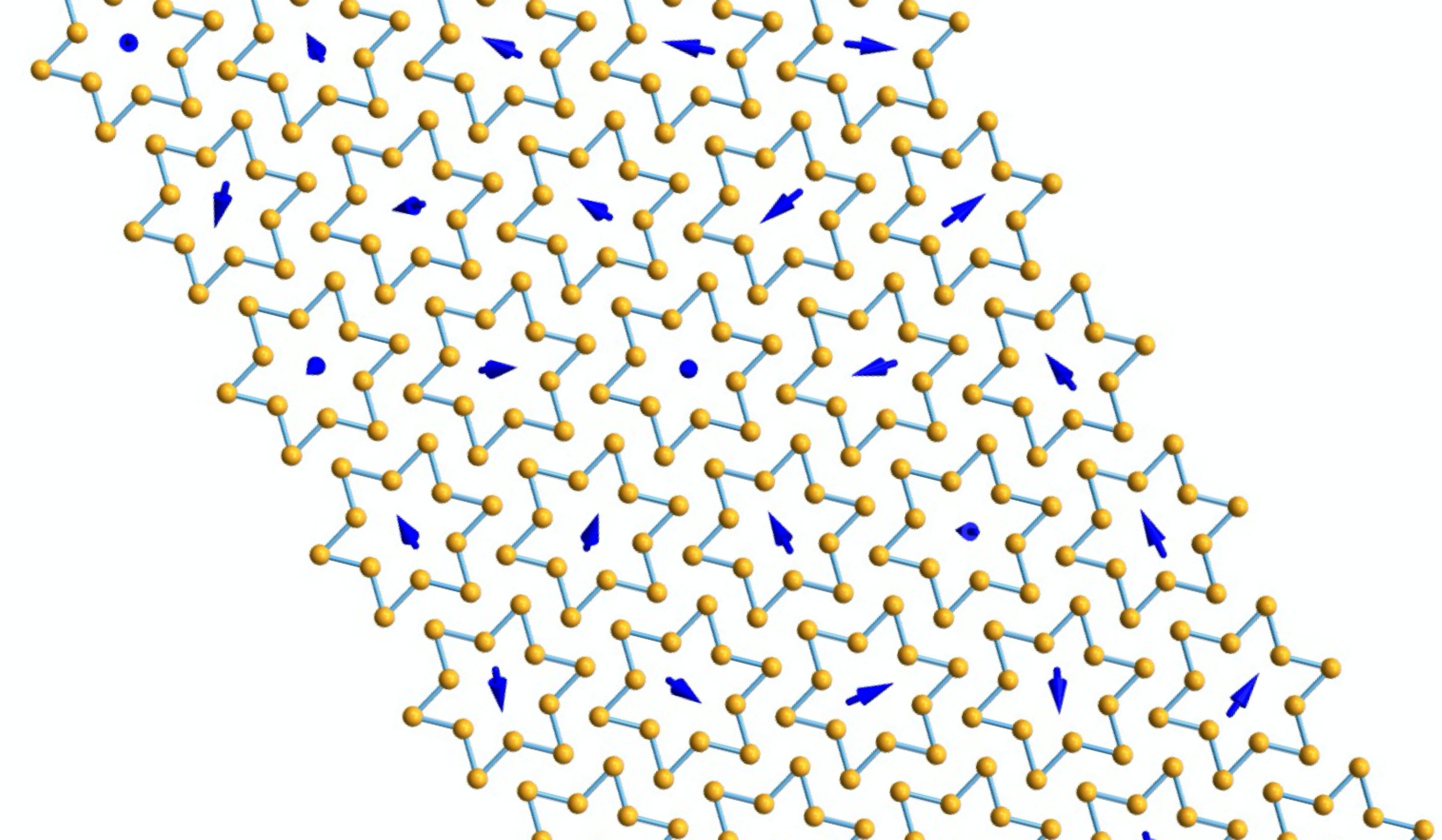}
	\caption{The super-triangular lattice formed by the star-of-David clusters on the 
	1T-TaS$_2$ in the commensurate CDW phase. Each star-of-David cluster traps one 
	unpaired spin-1/2 local moment. }
	\label{fig1}
\end{figure}

\noindent{\emph{\bf{2. Magnetic memory effect of superconducting 4Hb-TaS$_2$}}}\\
Due to various tunability and controllability of their physical properties, 
the TaS$_2$-based van der Waals materials~\cite{cryst7100298} 
have attracted a great interest. Among these TaS$_2$-based systems, 
the superconducting 4Hb-TaS$_2$ recently received some experimental 
attention~\cite{Persky2022,Shen_2022,Nayak_2021}. 
Differing from other TaS$_2$-based materials, the 4Hb-TaS$_2$ consists 
of an alternating stacking of 1T-TaS$_2$ and 1H-TaS$_2$ layers. 
Moreover, the octahedral 1T-TaS$_2$ and the trigonal prismatic 1H-TaS$_2$ 
layers have rather different physical properties. 
The multi-layered 1T-TaS$_2$ experiences a charge-density-wave (CDW) 
transition around 350K. At even lower temperatures below 200K, 
the system develops a commensurate CDW structure 
with a ${ \sqrt{13} \times \sqrt{13} }$ structure.  
The 1T-TaS$_2$ layer can then be viewed as a triangular lattice of clusters of stars of David (see Fig.~\ref{fig1}). 
In each cluster of star of David, there are 13 electrons and hence one unpaired electron. 
Due to the correlation, this unpaired electron forms an effective spin-1/2 local moment~\cite{Law_2017,PhysRevLett.129.017202,PhysRevLett.121.046401},
and the system becomes a correlation-driven cluster Mott insulator~\cite{PhysRevB.97.035124,PhysRevB.93.245134,PhysRevResearch.2.043424}. 
The emergence of the local spin-1/2 moment was supported by  
the Kondo resonance in the Pb-doped 1T-TaS$_2$~\cite{Shichao2022}. 
It was further shown that, this cluster Mott insulator is a spinon Fermi 
surface U(1) spin liquid in the weak Mott regime~\cite{PhysRevLett.121.046401},
and is quite analogous to the triangular lattice spin liquid in the organic compounds 
$\kappa$-(ET)$_2$Cu$_2$(CN)$_3$ and EtMe$_3$Sb[Pd(dmit)$_2$]$_2$~\cite{PhysRevLett.95.177001,PhysRevB.77.104413}. 
The muon-spin-relaxation ($\mu$SR) experiments indicated a gapless spin liquid 
dynamics and no long-range order down to 70mK in the 1T-TaS$_2$~\cite{Klanj_ek_2017}. 
The scanning tunneling microscopy on the isostructural   
monolayer 1T-TaSe$_2$ identified interesting evidences 
of the spinon Fermi surface in the cluster Mott insulating regime~\cite{Ruan_2021,Chen_2022}.  
In contrast to the correlated insulating behavior in the 1T-TaS$_2$, 
the 1H-TaS$_2$ layer develops the ${3 \times 3}$ CDW and 
is a good metal at low temperatures~\cite{Shen_2022}. 
The bulk form of 1H-TaS$_2$, 2H-TaS$_2$, 
superconducts at 0.7K. Once 1T-TaS$_2$ and 1H-TaS$_2$ 
layers are stacked together to form the bulk 4Hb-TaS$_2$ structure,
the superconducting transition temperature $T_c$ is greatly enhanced, reaching 2.7K. 
This anomalous enhancement might arise from 
the charge transfer between the 1T and the 1H layers~\cite{Persky2022,Shen_2022,Nayak_2021}. 
On the other hand, the insulating 1T-TaS$_2$ itself under the substitution of S with Se 
can superconduct with a maximal $T_c$ about 3.5K~\cite{PhysRevX.7.041054,2015NatCo...6.6091A}. 
Thus, it seems a bit unclear where the superconductivity originates from.
The more surprising and important result, however, is the      
magnetic memory effect and the spontaneous vortex generation,
and this is what we focus on below.  

% summary of spontaneous vortex generation 

We sketch the magnetic memory effect and the spontaneous vortex 
generation in 4Hb-TaS$_2$. It was found that, the system 
generates a vortex lattice under the field cooling from 
above $T_c$ to below $T_c$. After increasing the temperature 
above $T_c$ and turning off the magnetic field, one finds that  
a subsequent zero-field coupling to the temperatures below $T_c$ 
results in the superconducting vortices whose density is about 
one order of magnitude dilute than the field-cooling case. 
These vortices appeared spontaneously, 
suggesting an intrinsic magnetic memory of the system. 
This effect disappears when the thermal cycle
reaches 3.6K that is 0.9K above $T_c$. The magnetic signal 
was found to be absent both below $T_c$ (without the vortices) 
and above $T_c$~\cite{Persky2022}, which may be a bit incompatible 
with the chiral spin liquid (CSL) scenario as the scalar spin chirality 
would induce a weak magnetization.

% fate of the U(1) spin liquid, with superconducting pairing, 3D topological order, properties, vison line, phase transition , 

To understand the magnetic memory effect and the spontaneous vortex 
generation in 4Hb-TaS$_2$, we propose this physics is related to the $\mathbb{Z}_2$
topological order, and we expect the $\mathbb{Z}_2$ topological order to be relevant 
to the superconducting 4Hb-TaS$_2$. Here it does not really mean the $\mathbb{Z}_2$ 
topological order is the driving force for superconductivity, but simply suggests 
the potential relevance in this context.  
We start from the 1T-TaS$_2$ where the spinon Fermi surface U(1) spin liquid is suggested 
to be relevant~\cite{PhysRevLett.129.017202,PhysRevLett.121.046401,Ruan_2021}. 
The observation is that, once the electron pairing is introduced into the 
1T-TaS$_2$ system, the system becomes a $\mathbb{Z}_2$ spin liquid with a 
$\mathbb{Z}_2$ topological order. This is simply understood from a single-band
Hubbard model on a triangular lattice for the star of David clusters 
below,
\begin{eqnarray}
H = \sum_{\langle ij \rangle} ( -t c^{\dagger}_{i\sigma} c^{}_{j\sigma}  
 + \Delta c^{\dagger}_{i\uparrow} c^{\dagger}_{j\downarrow} + h.c.) 
 +  \sum_i U n_{i \uparrow} n_{i\downarrow} ,
\label{eq1}
\end{eqnarray}
where a uniform electron pairing is explicitly introduced. 
Here $c^{\dagger}_{i\sigma}$ ($c^{}_{i\sigma}$) is the creation
(annihilation) operator for the electron in the star-of-David cluster at the 
lattice site $i$, and ${n_{i\alpha} \equiv c^{\dagger}_{i\sigma} c^{}_{i\sigma}}$. 
Although the effective model for 1T-TaS$_2$ in the cluster Mott insulating regime 
was suggested to be a XXZ spin model with a XXZ-ring exchange term~\cite{PhysRevLett.121.046401,PhysRevB.72.045105}, 
the single-band Hubbard model is sufficient to capture the strong charge fluctuations 
for the emergence of the spin liquid and may be more convenient for the discussion 
of the pairing. The 1T-TaS$_2$ is located near the Mott transition~\cite{PhysRevLett.121.046401} 
and in the weak Mott insulating regime. 
Crudely speaking, this regime behaves like a metal at short distances, and is insulating at long distances.

This electron pairing may have several origins. It can arise from 
the electron-phonon coupling from either 1T-TaS$_2$ itself or the 1H layers. 
It can arise from the superconducting proximity effect from the 1H layers. 
It can also arise from the correlation effect in 1T-TaS$_2$ whose microscopic 
mechanism is less clear. In Ref.~\onlinecite{PhysRevLett.111.157203}, 
gapless $d$-wave pairing states both with and without time reversal symmetry
have actually been proposed. To capture the spin-charge separation 
in the weak Mott regime, we use the slave-rotor representation~\cite{PhysRevB.70.035114,PhysRevLett.95.036403} 
to express the electron operator as 
${c_{i\sigma} \equiv e^{-i\theta_i} f_{i\sigma}}$ where $e^{-i\theta_i}$ 
annihilates the boson charge at $i$ and $f_{i\sigma}$ is the annihilation
operator for the fermionic spinon, and the Hamiltonian without
the pairing becomes
\begin{eqnarray}
H_{\Delta=0} = \sum_{\langle ij \rangle} [ -t f^{\dagger}_{i\sigma} f^{}_{j\sigma}  e^{i \theta_i - i \theta_j }  + h.c.]
 +  \sum_i U n_{i \uparrow} n_{i\downarrow}.
\end{eqnarray}
The slave-rotor representation enlarges the physical Hilbert space, and 
a Hilbert space constraint ${L_i = \sum_{\sigma} f^{\dagger}_{i\sigma} f^{}_{i\sigma}}$ 
should be imposed where $L_i$ is a conjugate angular momentum operator for the rotor. 
With a standard slave-rotor mean-field decoupling~\cite{PhysRevLett.95.036403}, 
one finds the system enters the Mott side when ${U>2.73t}$. In the weak Mott regime,
the system in the absence of pairing is in a spinon Fermi surface U(1) spin liquid 
with the mean-field spinon Hamiltonian,
\begin{eqnarray}
H_{\text{U(1)}} = \sum_{\langle ij \rangle} [ -t_s f^{\dagger}_{i\sigma} f^{}_{j\sigma} + h.c.].
\end{eqnarray}
This mean-field result was equivalently established by working 
on the relevant exchange 
spin model for 1T-TaS$_2$ in the weak Mott regime~\cite{PhysRevLett.121.046401} 
and can be regarded as the parent state for the 1T-TaS$_2$ structure. 
The electron pairing term, in the Mott regime, becomes the spinon pairing 
in the mean-field description,
\begin{equation}
c^{\dagger}_{i\uparrow} c^{\dagger}_{j\downarrow} \rightarrow 
f^{\dagger}_{i\uparrow} f^{\dagger}_{j\downarrow} \langle e^{i \theta_i + i \theta_j } \rangle,
\end{equation}
and the U(1) gauge theory is immediately higgsed down to $\mathbb{Z}_2$. 
The resulting $\mathbb{Z}_2$ spin liquid is simply described by  
\begin{equation}
H_{\mathbb{Z}_2} = \sum_{ \langle ij \rangle } 
[- t_s  f^\dagger_{i \alpha} f^{}_{j \alpha}  
+ \Delta_s f^{\dagger}_{i\uparrow} f^{\dagger}_{j \downarrow} 
+ h.c.],
\label{eq3}
\end{equation}
where $t_s$ ($\Delta_s$) is the spinon hopping (pairing). We thus propose  
that the correlated 1T-TaS$_2$ layers in the superconducting environment 
are pertinent to the $\mathbb{Z}_2$ topological order. This is so as long as
there exist electron pairings in the 1T-TaS$_2$ layers, 
and we think this is the root to understand the exotic physics in the 4Hb-TaS$_2$. 
Depending on the strength of the interlayer coupling, this system can either be
a multi-layered two-dimensional $\mathbb{Z}_2$ topological order or 
be a three-dimensional $\mathbb{Z}_2$ topological order.

\begin{figure}[t] 
	\centering
	\includegraphics[width=6.8cm]{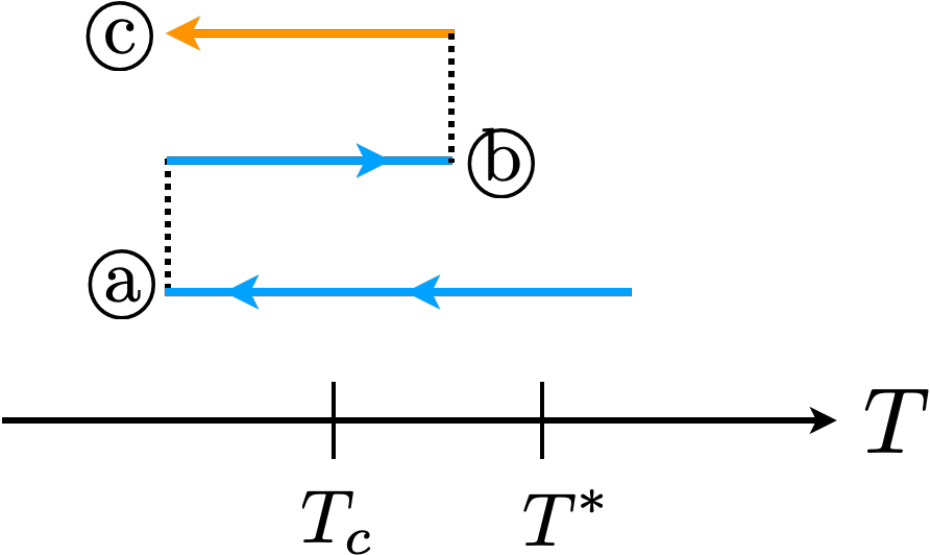}
	\caption{The thermal cycle for the vison-vortex nucleation. The arrow indicates the 
	time direction. The blue (orange) color represents the presence (absence) of the magnetic field. 
	The state at $\textcircled{a}$ is the field-driven vortex lattice. 
	The state at $\textcircled{b}$ is the vison crystal.
	The state at $\textcircled{c}$ is the spontaneously generated vortex lattice via the vison-vortex nucleation. 
	The dotted line is used for linkage and does not mean the system is fixed at this point.
	$T_c$ is the superconducting transition temperature. }
	\label{fig2}
\end{figure}

The magnetic memory effect and the spontaneous vortex generation 
in the 4Hb-TaS$_2$ can be understood from the proximity to the $\mathbb{Z}_2$
topological order. It is more convenient to discuss along the line of a three-dimensional 
$\mathbb{Z}_2$ topological order,
and the two-dimensional $\mathbb{Z}_2$ topological order requires a bit more 
explanation as it cannot exist at any finite temperature. 
We follow the thermal cycle in Fig.~\ref{fig2} (that is also the experimental cycle in Ref.~\onlinecite{Persky2022}) 
to describe the understanding from the $\mathbb{Z}_2$ topological order. 
The field cooling to the temperature below $T_c$ generates a large number of vortices 
in the form of vortex lattice (that corresponds to the state $\textcircled{a}$ in Fig.~\ref{fig2}). 
After raising the temperature above $T_c$, the magnetic field enters the system and the superconducting 
vortices just disappear. But the visons that are connected to the superconducting vortices
could remain in the system with the three-dimensional $\mathbb{Z}_2$ topological
order as long as the temperature is below the vison-unbinding transition at $T^{\ast}$.
The superconducting vortex only has one sign once the field direction is fixed. But 
the vison has a $\mathbb{Z}_2$ structure, and thus, a pair of visons can annihilate
each other. In the three-dimensional $\mathbb{Z}_2$ topological order, the visons 
appear as the vison loops or vison lines. This corresponds to the state $\textcircled{b}$. 
Before the vison lines come together and 
annihilate each other in the vison crystal, one cools the system back to the 
superconducting regime in the absence of the magnetic field. The vison crystal will 
nucleate back into the superconducting vortices (i.e. state $\textcircled{c}$ in Fig.~\ref{fig2}). 
This is a layered material, the nucleated 
vortices have a natural orientation. The sign of the flux for the vortices 
is selected by the residual magnetic field from the environment. 
This spontaneous vortex generation will disappear if the temperature is raised 
above $T^{\ast}$ or the time internal for the system above $T_c$ 
is too long such that the visons have already annihilated completely. 
If one performs the experiment with a single or odd number of vortices, 
the consequence of this annihilation will always end up with one residual vison, 
and thus, one would always obtain a single nucleated vortice. 
In the discussion part, we mention the modification with the multi-layered
two-dimensional $\mathbb{Z}_2$ topological order where the vison-vortex 
nucleation idea stays invariant. 
 \\

% vortex-vison nucleation 
% even vs odd 

\begin{figure}[t] 
	\centering
	\includegraphics[width=7.2cm]{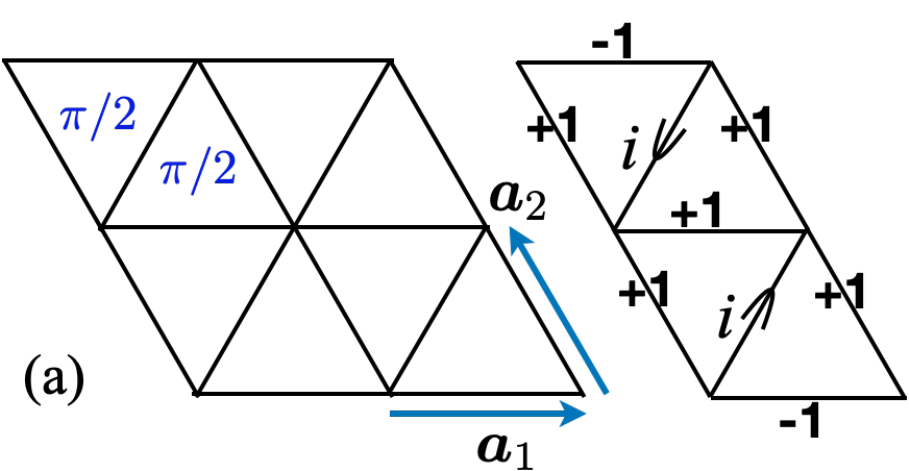}
	\includegraphics[width=8.6cm]{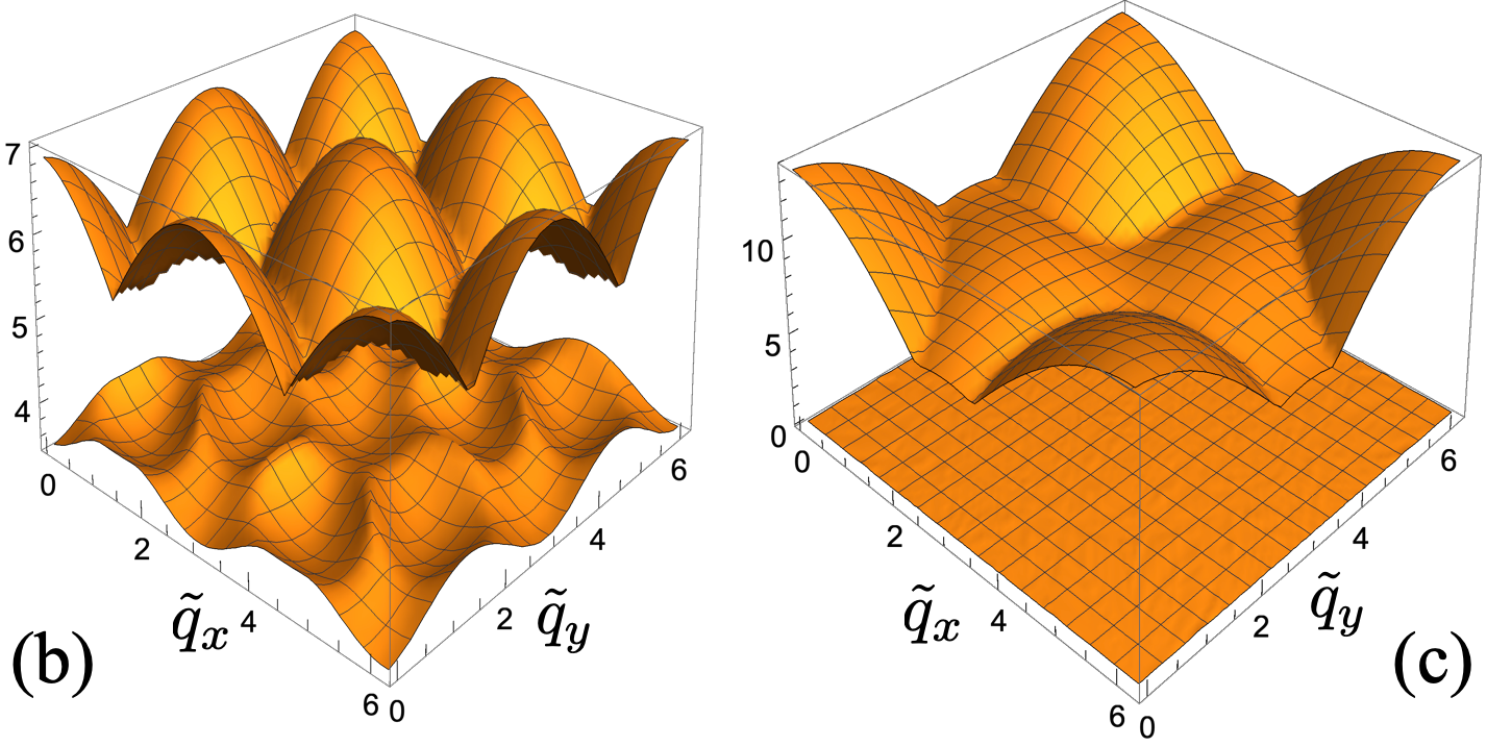}
	\caption{(a) The gauge flux pattern in the CSL and its gauge fixing (that is repeated along ${\boldsymbol a}_1$ direction).  
	The length of the lattice vectors is set to unity.
	The upper and lower excitation edges of the spinon continuum for the CSL in (b)
    and for the $\mathbb{Z}_2$ topological order of Eq.~\eqref{eq3} in (c). 
	The energy unit is set to $t_s$, and ${\Delta_s = 0.2t_s}$. For our convenient, we have defined 
	${\tilde{q}_x \equiv {\boldsymbol q}\cdot {\boldsymbol a}_1}$ and ${\tilde{q}_y \equiv {\boldsymbol q}\cdot {\boldsymbol a}_2}$.
	 }
	\label{fig3}
\end{figure}

\noindent{\emph{\bf 3. Distinction from chiral spin liquid scenario}}\\
We suggest further experiments to distinguish the proposal with the 
$\mathbb{Z}_2$ topological order from the proposal with the chiral spin liquid (CSL). 
The CSL was proposed numerically for the 
triangular lattice Hubbard model in the weak Mott regime and was theoretically 
analyzed~\cite{PhysRevX.10.021042,PhysRevLett.127.087201}. 
The time reversal symmetry breaking with the scalar spin chirality was shown to arise
from the ring exchange interaction that can be thought as the products 
of the scalar spin chiralities~\cite{PhysRevLett.127.087201}. 
Due to the connection between the 1T-TaS$_2$ 
and the triangular lattice Hubbard model and/or the ring exchange model, 
such a CSL may have a relevance with the superconducting 4Hb-TaS$_2$. 
As the CSL breaks time reversal symmetry and time reversal symmetry 
is an Ising symmetry, there should be a finite temperature transition.  
In addition, there exists a uniform distribution of the scalar spin chirality 
throughout the triangular lattice that corresponds to the emergent 
U(1) gauge flux distribution with a $\pi/2$ flux on each triangle 
and $\pi$ flux on each unit cell. In the CSL, the spinon 
is already fully gapped. Moreover, due to the presence of the background 
$\pi$ flux, the translation symmetry is realized projectively for the spinons,
and the spinon continuum that is detectable by the inelastic neutron scattering 
experiments would have an enhanced spectral periodicity in the Brillouin 
zone~\cite{PhysRevB.96.085136,PhysRevB.96.195127,PhysRevB.90.121102,PhysRevB.87.104406}. 
To reveal this, we combine a generic argument~\cite{PhysRevB.90.121102} 
with the calculation by fixing the gauge according to Fig.~\ref{fig3}.     
\\

\noindent{\emph{\bf 3.1 Spectral periodicity of spinon continuum in chiral spin liquid}}\\
One first defines the two lattice translation operations $T_1$ and $T_2$, where 
the $T_1$ ($T_2$) operation translates the system by the triangular lattice 
vector ${\boldsymbol a}_1$ (${\boldsymbol a}_2$). According to Fig.3 in the main text.
we have 
\begin{eqnarray}
{\boldsymbol a}_1 &=& (1,0) , \\
{\boldsymbol a}_2 &=& (-\frac{1}{2}, \frac{\sqrt{3}}{2}) .
\end{eqnarray}

For the chiral spin liquid, the spinons are defined and fractionalized excitations, in addition
to the fractional statistics. The symmetry is again fractionalized , and more precisely, the symmetry
operation acts locally the spinons. For the translation symmetry, if we translate the spinon
around the parallelogram formed by ${\boldsymbol a}_1$ and ${\boldsymbol a}_2$, the spinon 
will experience the background $\pi$ flux~\cite{PhysRevB.65.165113,PhysRevB.90.121102,PhysRevB.96.085136}, i.e.
\begin{eqnarray}
(T_2^s)^{-1} (T_1^s)^{-1} T_2^s \, T_1^s = -1 ,
\label{seq1}
\end{eqnarray}
where the superscript `s' merely refers to the spinon symmetry operation. Thus, for the spinon,
these two translation operations anticommute with each other with,
\begin{equation}
 T_1^s   T_2^s = - T_2^s  T_1^s . 
 \label{seq2}
\end{equation}

The translation symmetry fractionalization and the anticommutation relation between the two 
translation operations immediately lead to an enhanced spectral periodicity 
of the spinon continuum in the chiral spin liquid. 
To show that, we extend the $\mathbb{Z}_2$ topological order of 
the square lattice example in Ref.~\onlinecite{PhysRevB.90.121102} to 
the triangular lattice and consider a generic two-spinon scattering state, 
\begin{equation}
| a \rangle \equiv |  {\boldsymbol q}_a , m_a \rangle 
\end{equation}
where ${\boldsymbol q}_a$ refers to the crystal momentum of this state $| a \rangle $,
and $m_a$ labels the rest quantum number to characterize this state. 
Because the Bravais lattice vectors of the triangular lattice are 
not orthogonal, we expand the crystal momentum ${\boldsymbol q}_a$ 
in the nonorthogonal basis in the reciprocal space with
\begin{equation}
{\boldsymbol q}_a = q_{a1} {\boldsymbol e}_1 + q_{a2} {\boldsymbol e}_2,
\end{equation}
where the basis vectors are given as 
\begin{eqnarray}
{\boldsymbol e}_1 & = & (0,\frac{2}{\sqrt{3}}) ,\\
{\boldsymbol e}_2 & = & (1, \frac{1}{\sqrt{3}}) . 
\end{eqnarray}
Based on the symmetry fractionalization and the symmetry localization, we apply the 
lattice translation on the two-spinon scattering state and obtain, 
\begin{eqnarray}
T_{\mu}  | a \rangle = T^s_{\mu} ( s1)  T^s_{\mu} ( s2)    | a \rangle.
\end{eqnarray}
 Here `$s1$' and `$s2$' refer to two spinons, and the translation operation $T_{\mu}$
 is fractionalized to two spinon operations $T_{\mu}^s$. 
 We then apply the spinon translation operation on the spinon $s1$ of the two-spinon
 scattering state $| a \rangle$ to generate other two equal energy spinon scattering 
 states, i.e.,
 \begin{eqnarray}
 |b \rangle &=&  T_1^s (s1) |a\rangle, \\
 |c \rangle &=&  T_2^s (s1) |a\rangle. 
 \end{eqnarray}
To show that these spinon scattering states have distinct crystal momenta, we 
apply the lattice translation operation on these states and compare the eigenvalues,
\begin{eqnarray}
T_1 |b \rangle  &=& T_1^s (s1) T_1^s (s2) T_1^s (s1) |a\rangle = + T_1^s (s1) [T_1 |a\rangle] ,
\label{seq22}
\\
T_2 |b \rangle &=&  T_2^s (s1) T_2^s (s2) T_1^s (s1) |a\rangle = - T_1^s (s1) [T_2 |a \rangle].
\label{seq23}
\end{eqnarray}
In Eq.~\eqref{seq23}, the anticommutation relation in Eq.~\eqref{seq2} has been used. 
The above relations indicates that, 
\begin{eqnarray}
q_{b1} = q_{a1}, \quad \quad q_{b2} = q_{a2} + \pi. 
\end{eqnarray} 

With a similar construction, we find that, 
\begin{eqnarray}
T_1 |c \rangle &=& -T_2^s (s1) [ T_1 |a\rangle ] ,
\\
T_2 |c \rangle &=& + T_2^s (s1) [ T_2 |a\rangle ] .  
\end{eqnarray}
Thus, we have 
\begin{eqnarray}
q_{c1} = q_{a1} + \pi , \quad\quad q_{c2}=q_{a2}. 
\end{eqnarray}
These two-spinon scattering states have the same energy, and the relation between
their crystal momenta indicates that there is an enhanced spectral periodicity 
in the spinon continuum. Both the lower and upper excitation edges of the spinon continuum
have this spectral periodicity enhancement with
\begin{eqnarray}
&& {\mathcal L} [{\boldsymbol q}] = {\mathcal L} [{\boldsymbol q} + \pi {\boldsymbol e}_1] = {\mathcal L} [{\boldsymbol q} + \pi {\boldsymbol e}_2] , \\
&& {\mathcal U} [{\boldsymbol q}] = {\mathcal U} [{\boldsymbol q} + \pi {\boldsymbol e}_1 ]= {\mathcal U} [{\boldsymbol q} + \pi {\boldsymbol e}_2]. 
\end{eqnarray}
An explicit calculation of the spinon continuum for the chiral spin liquid is given in the following.
\\ 

%%%

\noindent{\emph{\bf 3.2 Distinction in spinon continuum}}\\
As we have discussed in details above, both the lower excitation edge,  
$\mathcal{L} [{\boldsymbol q}]$, and the upper excitation edge, 
$\mathcal{U} [{\boldsymbol q}]$, of the spinon continuum develop an enhanced spectral periodicity.  
These two edges are the energy bounds for the spinon continuum, and 
we have 
\begin{eqnarray}
&& \mathcal{L} [{\boldsymbol q}]= \mathcal{L} [{\boldsymbol q} + \pi (0,\frac{2}{\sqrt{3}})] = \mathcal{L} [{\boldsymbol q} + \pi (1,\frac{1}{\sqrt{3}})],
\\
&& \mathcal{U} [{\boldsymbol q}]= \mathcal{U} [{\boldsymbol q} + \pi (0,\frac{2}{\sqrt{3}})] = \mathcal{U} [{\boldsymbol q} + \pi (1,\frac{1}{\sqrt{3}})]. 
\end{eqnarray}
For the explicit calculation, we consider the following mean-field spinon Hamiltonian for the two-dimensional CSL,
\begin{equation}
H_{\text{CSL}} = 
\sum_{\langle ij \rangle} 
\big[ 
{-t_s e^{i \phi_{ij}} f^\dagger_{i \alpha} f^{}_{j \alpha} + h.c.  } 
\big] ,
\end{equation}
where the phase $e^{i \phi_{ij}}$ is chosen according to Fig.~\ref{fig3}. 
The spinons have two bands and the dispersions are given as
\begin{eqnarray}
\omega_{\pm} ({\boldsymbol q}) &=& \pm t_s \big[2 \big(3 + \cos ( {2 {\boldsymbol q}\cdot {\boldsymbol a}_1 } ) 
+  \cos ( {2{\boldsymbol q}\cdot {\boldsymbol a}_2 } )
\nonumber \\
&& \quad\quad \quad \quad +  \cos ( {2{\boldsymbol q}\cdot {\boldsymbol a}_1 +2{\boldsymbol q}\cdot {\boldsymbol a}_2 } ) \big)\big]^{1/2}.
\end{eqnarray} 
The lowest spinon band carries a Chern number ${C=1}$, and is fully filled. 
The gapped spinon continuum $\Omega ({\boldsymbol q})$ is obtained from 
${{\boldsymbol q} = {\boldsymbol q}_1 - {\boldsymbol q}_2}$ 
and ${\Omega ({\boldsymbol q}) = \omega_+({\boldsymbol q}_1) 
- \omega_-({\boldsymbol q}_2)}$, and is plotted in Fig.~\ref{fig3}. 

As a comparison, we compute the spinon continuum for the $\mathbb{Z}_2$ topological order on the triangular lattice
from Eq.~\eqref{eq3}.
As this state is derived from the spinon Fermi surface state with the 
uniform hopping by introducing the pairing, we do not expect the spectral 
periodicity enhancement (see Fig.~\ref{fig3}). Nevertheless, due to the 
weak uniform pairing, the spectral weight of the weakly gapped spinon 
continuum may still carry the ``$2k_F$'' features of the spinon Fermi surface~\cite{PhysRevLett.121.046401}. 
\\

\noindent{\emph{\bf 3.3 Other distinctions}}\\
In addition to the distinction between the spectral periodicity in the spinon continuum, 
the thermal Hall effect is another clear experimental probe to distinguish 
the $\mathbb{Z}_2$ topological order from the CSL. 
Due to the finite Chern number of the filled spinon band in the CSL, 
there exists a chiral (charge-neutral) edge state that does not really conduct
charge but conduct heat. One expects a quantized thermal Hall effect with 
\begin{eqnarray}
{\kappa_{xy}}/{T}= \frac{\pi}{3} k_B^2 / \hbar ,
\end{eqnarray}
even in the absence of magnetic field, and the gauge fluctuations do not
change the order of magnitude of $\kappa_{xy}/T$~\cite{PhysRevB.101.195126}.
To avoid the complication from the superconducting vortices, one may carry
the measurement above $T_c$. This quantized thermal Hall effect should be 
a large signal if the CSL is relevant for the 4Hb-TaS$_2$. 
Since there exists the charge transfer between the layers in 4Hb-TaS$_2$, 
one further expects a topological Hall effect from the charge carriers, where the 
charge carriers experience the scalar spin chirality from the CSL as the real-space Berry
phase and produce the Hall effect. 
In contrast, both anomalous Hall effect and thermal Hall effect should be absent in
our proposal for the $\mathbb{Z}_2$ topological order, 
except if the superconducting phase has a weak time reversal symmetry
breaking~\cite{sciadv.aax9480,PhysRevLett.111.157203},
and the signals may be quite weak compared to the quantized
one in the CSL scenario.  
\\
 
 % topological Hall effect 
 
% symmetry fractionalization 

\noindent{\emph{\bf 4. Discussion}}\\
We compare the vison-vortex nucleation idea with the vortex lattice 
here and the single vortex case in the Senthil-Fisher cuprate proposal~\cite{PhysRevLett.86.292}. 
In the cuprate proposal, Senthil and Fisher considered a single $hc/2e$ 
vortex. The system leaves the superconducting regime, the single vison 
puts the system in a distinct superselection sector of the 
$\mathbb{Z}_2$ topological order, and the single vison cannot annihilate 
itself. Once the system reenters the superconducting regime, 
the vison would nucleate into a superconducting vortices. They further argued that, 
as long as the number of vortices is odd, there will be one vison remaining 
and the phenomenon will be there. For the case of the vortex lattice that we discuss in this Letter, 
once the system leaves the superconducting regime but remains in the 
${\mathbb Z}_2$ topologically ordered regime, the visons in the vison crystal would 
annihilate with each other. But if this time for the vison crystal to be fully 
annihilated is longer than the time interval for the experiments to return to the superconducting regime, 
the remaining visons will nucleate back to the superconducting vortices.
This seems to be what has happened in the superconducting 4Hb-TaS$_2$~\cite{Persky2022}.
Thus the vortex lattice with many vortices makes the experimental phenomena 
much more visible than a single vortex. 
 
The three-dimensional $\mathbb{Z}_2$ topological order has a finite temperature
phase transition from the vison loop unbinding. This transition
is an Ising transition, and sets the upper temperature for 
the magnetic memory effect. Indeed, in the low-temperature thermodynamic measurement 
in the 4Hb-TaS$_2$, only the superconducting transition is clearly visible~\cite{sciadv.aax9480}, 
and the upper temperature for the magnetic memory effect is 3.6K, a bit above $T_c$. 
Due to the layered structure, it is also possible, the $\mathbb{Z}_2$ topological order
is two dimensional. In such a case, strictly speaking, the topological order cannot exist
at any finite temperature. Similar to what has been argued by Senthil and Fisher for the single vison,
the vison crystal, however, can be present for a finite time before escaping.
In this case, besides the vison annihilation effect, the vison gap 
sets a crossover temperature scale for the upper temperature for 
the magnetic memory effect. 

For the chiral spin liquid scenario, one may invoke the anyon superconductivity.
As the anyon naturally carries the intrinsic flux,  one may establish 
the connection between the anyon excitations and the superconducting vortices in a way
analogous to the vison and the vortex. In 
the state $\textcircled{b}$ in Fig.~\ref{fig2}, one would expect an anyon crystal of some sort. 
We will come back to this theoretical possibility in the future work. 
 
Since we expect the $\mathbb{Z}_2$ topological order arises from the correlation 
physics of the 1T-TaS$_2$ layers, it is natural for us to vision that a similar magnetic 
magnetic memory effect and the spontaneous vortex generation could occur
in the Se-doped 1T-TaS$_2$ where the superconductivity is also obtained~\cite{PhysRevX.7.041054,2015NatCo...6.6091A}. 

To summarize, we associate the puzzling magnetic memory effect 
and the spontaneous vortex generation in the  
4Hb-TaS$_2$ to the vison-vortex nucleation from the $\mathbb{Z}_2$ topological order, 
and suggest other experimental phenomena to distinguish this proposal from 
the chiral spin liquid scenario. 
\\

\noindent{\emph{\bf Acknowledgments}}\\
We thank Shichao Yan, Yi Yin, Yuan Li, Kui Jin, Beena Kalisky and especially Matthew Fisher for insightful discussion.  
\\

\noindent{\emph{\bf Funding}}\\
Open access funding provided by Shanghai Jiao Tong University. 
This work is supported by 
the Ministry of Science and Technology of China with Grants No.~2021YFA1400300,
and the National Science Foundation of China with Grant No.~92065203.
\\

\noindent{\emph{\bf Availability of data and materials}}\\
All data and materials are included in the main text. 
Further information can be requested from corresponding author via email.
\\

\noindent{\emph{\bf Declarations}}\\

\noindent{\emph{\bf Competing interests}}\\
Gang V. Chen is an editorial board member for Quantum Frontiers and was not involved in the editorial review, or the decision to publish, this article. 
Both authors declare that there are no competing interests.
\\

\noindent{\emph{\bf Author contributions}}\\
GC designed and supervised this project. Rui Luo was involved in the discussion and the further development
of this project. All authors commented on the results.

\bibliography{RefsJuly.bib}

\begin{thebibliography}{44}%
\makeatletter
\providecommand \@ifxundefined [1]{%
 \@ifx{#1\undefined}
}%
\providecommand \@ifnum [1]{%
 \ifnum #1\expandafter \@firstoftwo
 \else \expandafter \@secondoftwo
 \fi
}%
\providecommand \@ifx [1]{%
 \ifx #1\expandafter \@firstoftwo
 \else \expandafter \@secondoftwo
 \fi
}%
\providecommand \natexlab [1]{#1}%
\providecommand \enquote  [1]{``#1''}%
\providecommand \bibnamefont  [1]{#1}%
\providecommand \bibfnamefont [1]{#1}%
\providecommand \citenamefont [1]{#1}%
\providecommand \href@noop [0]{\@secondoftwo}%
\providecommand \href [0]{\begingroup \@sanitize@url \@href}%
\providecommand \@href[1]{\@@startlink{#1}\@@href}%
\providecommand \@@href[1]{\endgroup#1\@@endlink}%
\providecommand \@sanitize@url [0]{\catcode `\\12\catcode `\$12\catcode
  `\&12\catcode `\#12\catcode `\^12\catcode `\_12\catcode `\%12\relax}%
\providecommand \@@startlink[1]{}%
\providecommand \@@endlink[0]{}%
\providecommand \url  [0]{\begingroup\@sanitize@url \@url }%
\providecommand \@url [1]{\endgroup\@href {#1}{\urlprefix }}%
\providecommand \urlprefix  [0]{URL }%
\providecommand \Eprint [0]{\href }%
\providecommand \doibase [0]{http://dx.doi.org/}%
\providecommand \selectlanguage [0]{\@gobble}%
\providecommand \bibinfo  [0]{\@secondoftwo}%
\providecommand \bibfield  [0]{\@secondoftwo}%
\providecommand \translation [1]{[#1]}%
\providecommand \BibitemOpen [0]{}%
\providecommand \bibitemStop [0]{}%
\providecommand \bibitemNoStop [0]{.\EOS\space}%
\providecommand \EOS [0]{\spacefactor3000\relax}%
\providecommand \BibitemShut  [1]{\csname bibitem#1\endcsname}%
\let\auto@bib@innerbib\@empty
%</preamble>
\bibitem [{\citenamefont {Wen}(2013)}]{Wen_2013}%
  \BibitemOpen
  \bibfield  {author} {\bibinfo {author} {\bibfnamefont {Xiao-Gang}\
  \bibnamefont {Wen}},\ }\bibfield  {title} {\enquote {\bibinfo {title}
  {{Topological Order: From Long-Range Entangled Quantum Matter to a Unified
  Origin of Light and Electrons}},}\ }\href {\doibase 10.1155/2013/198710}
  {\bibfield  {journal} {\bibinfo  {journal} {{ISRN} Condensed Matter Physics}\
  }\textbf {\bibinfo {volume} {2013}},\ \bibinfo {pages} {1--20} (\bibinfo
  {year} {2013})}\BibitemShut {NoStop}%
\bibitem [{\citenamefont {Wen}(2002)}]{PhysRevB.65.165113}%
  \BibitemOpen
  \bibfield  {author} {\bibinfo {author} {\bibfnamefont {Xiao-Gang}\
  \bibnamefont {Wen}},\ }\bibfield  {title} {\enquote {\bibinfo {title}
  {Quantum orders and symmetric spin liquids},}\ }\href {\doibase
  10.1103/PhysRevB.65.165113} {\bibfield  {journal} {\bibinfo  {journal} {Phys.
  Rev. B}\ }\textbf {\bibinfo {volume} {65}},\ \bibinfo {pages} {165113}
  (\bibinfo {year} {2002})}\BibitemShut {NoStop}%
\bibitem [{\citenamefont {Mesaros}\ and\ \citenamefont
  {Ran}(2013)}]{PhysRevB.87.155115}%
  \BibitemOpen
  \bibfield  {author} {\bibinfo {author} {\bibfnamefont {Andrej}\ \bibnamefont
  {Mesaros}}\ and\ \bibinfo {author} {\bibfnamefont {Ying}\ \bibnamefont
  {Ran}},\ }\bibfield  {title} {\enquote {\bibinfo {title} {Classification of
  symmetry enriched topological phases with exactly solvable models},}\ }\href
  {\doibase 10.1103/PhysRevB.87.155115} {\bibfield  {journal} {\bibinfo
  {journal} {Phys. Rev. B}\ }\textbf {\bibinfo {volume} {87}},\ \bibinfo
  {pages} {155115} (\bibinfo {year} {2013})}\BibitemShut {NoStop}%
\bibitem [{\citenamefont {Essin}\ and\ \citenamefont
  {Hermele}(2013)}]{PhysRevB.87.104406}%
  \BibitemOpen
  \bibfield  {author} {\bibinfo {author} {\bibfnamefont {Andrew~M.}\
  \bibnamefont {Essin}}\ and\ \bibinfo {author} {\bibfnamefont {Michael}\
  \bibnamefont {Hermele}},\ }\bibfield  {title} {\enquote {\bibinfo {title}
  {{Classifying fractionalization: Symmetry classification of gapped
  ${\mathbb{Z}}_{2}$ spin liquids in two dimensions}},}\ }\href {\doibase
  10.1103/PhysRevB.87.104406} {\bibfield  {journal} {\bibinfo  {journal} {Phys.
  Rev. B}\ }\textbf {\bibinfo {volume} {87}},\ \bibinfo {pages} {104406}
  (\bibinfo {year} {2013})}\BibitemShut {NoStop}%
\bibitem [{\citenamefont {Kitaev}(2006)}]{Kitaev_2006}%
  \BibitemOpen
  \bibfield  {author} {\bibinfo {author} {\bibfnamefont {Alexei}\ \bibnamefont
  {Kitaev}},\ }\bibfield  {title} {\enquote {\bibinfo {title} {Anyons in an
  exactly solved model and beyond},}\ }\href {\doibase
  10.1016/j.aop.2005.10.005} {\bibfield  {journal} {\bibinfo  {journal} {Annals
  of Physics}\ }\textbf {\bibinfo {volume} {321}},\ \bibinfo {pages} {2--111}
  (\bibinfo {year} {2006})}\BibitemShut {NoStop}%
\bibitem [{\citenamefont {Laughlin}(1983)}]{PhysRevLett.50.1395}%
  \BibitemOpen
  \bibfield  {author} {\bibinfo {author} {\bibfnamefont {R.~B.}\ \bibnamefont
  {Laughlin}},\ }\bibfield  {title} {\enquote {\bibinfo {title} {{Anomalous
  Quantum Hall Effect: An Incompressible Quantum Fluid with Fractionally
  Charged Excitations}},}\ }\href {\doibase 10.1103/PhysRevLett.50.1395}
  {\bibfield  {journal} {\bibinfo  {journal} {Phys. Rev. Lett.}\ }\textbf
  {\bibinfo {volume} {50}},\ \bibinfo {pages} {1395--1398} (\bibinfo {year}
  {1983})}\BibitemShut {NoStop}%
\bibitem [{\citenamefont {Arovas}\ \emph {et~al.}(1984)\citenamefont {Arovas},
  \citenamefont {Schrieffer},\ and\ \citenamefont
  {Wilczek}}]{PhysRevLett.53.722}%
  \BibitemOpen
  \bibfield  {author} {\bibinfo {author} {\bibfnamefont {Daniel}\ \bibnamefont
  {Arovas}}, \bibinfo {author} {\bibfnamefont {J.~R.}\ \bibnamefont
  {Schrieffer}}, \ and\ \bibinfo {author} {\bibfnamefont {Frank}\ \bibnamefont
  {Wilczek}},\ }\bibfield  {title} {\enquote {\bibinfo {title} {{Fractional
  Statistics and the Quantum Hall Effect}},}\ }\href {\doibase
  10.1103/PhysRevLett.53.722} {\bibfield  {journal} {\bibinfo  {journal} {Phys.
  Rev. Lett.}\ }\textbf {\bibinfo {volume} {53}},\ \bibinfo {pages} {722--723}
  (\bibinfo {year} {1984})}\BibitemShut {NoStop}%
\bibitem [{\citenamefont {Senthil}\ and\ \citenamefont
  {Fisher}(2000)}]{PhysRevB.62.7850}%
  \BibitemOpen
  \bibfield  {author} {\bibinfo {author} {\bibfnamefont {T.}~\bibnamefont
  {Senthil}}\ and\ \bibinfo {author} {\bibfnamefont {Matthew P.~A.}\
  \bibnamefont {Fisher}},\ }\bibfield  {title} {\enquote {\bibinfo {title}
  {{${Z}_{2}$ gauge theory of electron fractionalization in strongly correlated
  systems}},}\ }\href {\doibase 10.1103/PhysRevB.62.7850} {\bibfield  {journal}
  {\bibinfo  {journal} {Phys. Rev. B}\ }\textbf {\bibinfo {volume} {62}},\
  \bibinfo {pages} {7850--7881} (\bibinfo {year} {2000})}\BibitemShut {NoStop}%
\bibitem [{\citenamefont {Senthil}\ and\ \citenamefont
  {Fisher}(2001)}]{PhysRevLett.86.292}%
  \BibitemOpen
  \bibfield  {author} {\bibinfo {author} {\bibfnamefont {T.}~\bibnamefont
  {Senthil}}\ and\ \bibinfo {author} {\bibfnamefont {Matthew P.~A.}\
  \bibnamefont {Fisher}},\ }\bibfield  {title} {\enquote {\bibinfo {title}
  {{Fractionalization in the Cuprates: Detecting the Topological Order}},}\
  }\href {\doibase 10.1103/PhysRevLett.86.292} {\bibfield  {journal} {\bibinfo
  {journal} {Phys. Rev. Lett.}\ }\textbf {\bibinfo {volume} {86}},\ \bibinfo
  {pages} {292--295} (\bibinfo {year} {2001})}\BibitemShut {NoStop}%
\bibitem [{\citenamefont {Balents}\ \emph {et~al.}(2000)\citenamefont
  {Balents}, \citenamefont {Fisher},\ and\ \citenamefont
  {Nayak}}]{PhysRevB.61.6307}%
  \BibitemOpen
  \bibfield  {author} {\bibinfo {author} {\bibfnamefont {Leon}\ \bibnamefont
  {Balents}}, \bibinfo {author} {\bibfnamefont {Matthew P.~A.}\ \bibnamefont
  {Fisher}}, \ and\ \bibinfo {author} {\bibfnamefont {Chetan}\ \bibnamefont
  {Nayak}},\ }\bibfield  {title} {\enquote {\bibinfo {title} {Dual vortex
  theory of strongly interacting electrons: A non-fermi liquid with a twist},}\
  }\href {\doibase 10.1103/PhysRevB.61.6307} {\bibfield  {journal} {\bibinfo
  {journal} {Phys. Rev. B}\ }\textbf {\bibinfo {volume} {61}},\ \bibinfo
  {pages} {6307--6319} (\bibinfo {year} {2000})}\BibitemShut {NoStop}%
\bibitem [{\citenamefont {Balents}\ \emph {et~al.}(1999)\citenamefont
  {Balents}, \citenamefont {Fisher},\ and\ \citenamefont
  {Nayak}}]{PhysRevB.60.1654}%
  \BibitemOpen
  \bibfield  {author} {\bibinfo {author} {\bibfnamefont {Leon}\ \bibnamefont
  {Balents}}, \bibinfo {author} {\bibfnamefont {Matthew P.~A.}\ \bibnamefont
  {Fisher}}, \ and\ \bibinfo {author} {\bibfnamefont {Chetan}\ \bibnamefont
  {Nayak}},\ }\bibfield  {title} {\enquote {\bibinfo {title} {Dual order
  parameter for the nodal liquid},}\ }\href {\doibase 10.1103/PhysRevB.60.1654}
  {\bibfield  {journal} {\bibinfo  {journal} {Phys. Rev. B}\ }\textbf {\bibinfo
  {volume} {60}},\ \bibinfo {pages} {1654--1667} (\bibinfo {year}
  {1999})}\BibitemShut {NoStop}%
\bibitem [{\citenamefont {Motrunich}\ and\ \citenamefont
  {Senthil}(2005)}]{PhysRevB.71.125102}%
  \BibitemOpen
  \bibfield  {author} {\bibinfo {author} {\bibfnamefont {O.~I.}\ \bibnamefont
  {Motrunich}}\ and\ \bibinfo {author} {\bibfnamefont {T.}~\bibnamefont
  {Senthil}},\ }\bibfield  {title} {\enquote {\bibinfo {title} {Origin of
  artificial electrodynamics in three-dimensional bosonic models},}\ }\href
  {\doibase 10.1103/PhysRevB.71.125102} {\bibfield  {journal} {\bibinfo
  {journal} {Phys. Rev. B}\ }\textbf {\bibinfo {volume} {71}},\ \bibinfo
  {pages} {125102} (\bibinfo {year} {2005})}\BibitemShut {NoStop}%
\bibitem [{\citenamefont {Senthil}(2015)}]{Senthil_2015}%
  \BibitemOpen
  \bibfield  {author} {\bibinfo {author} {\bibfnamefont {T.}~\bibnamefont
  {Senthil}},\ }\bibfield  {title} {\enquote {\bibinfo {title}
  {Symmetry-protected topological phases of quantum matter},}\ }\href {\doibase
  10.1146/annurev-conmatphys-031214-014740} {\bibfield  {journal} {\bibinfo
  {journal} {Annual Review of Condensed Matter Physics}\ }\textbf {\bibinfo
  {volume} {6}},\ \bibinfo {pages} {299--324} (\bibinfo {year}
  {2015})}\BibitemShut {NoStop}%
\bibitem [{\citenamefont {Chen}(2016)}]{PhysRevB.94.205107}%
  \BibitemOpen
  \bibfield  {author} {\bibinfo {author} {\bibfnamefont {Gang}\ \bibnamefont
  {Chen}},\ }\bibfield  {title} {\enquote {\bibinfo {title} {{``Magnetic
  monopole'' condensation of the pyrochlore ice U(1) quantum spin liquid:
  Application to ${\mathrm{Pr}}_{2}{\mathrm{Ir}}_{2}{\mathrm{O}}_{7}$ and
  ${\mathrm{Yb}}_{2}{\mathrm{Ti}}_{2}{\mathrm{O}}_{7}$}},}\ }\href {\doibase
  10.1103/PhysRevB.94.205107} {\bibfield  {journal} {\bibinfo  {journal} {Phys.
  Rev. B}\ }\textbf {\bibinfo {volume} {94}},\ \bibinfo {pages} {205107}
  (\bibinfo {year} {2016})}\BibitemShut {NoStop}%
\bibitem [{\citenamefont {Bonn}\ \emph {et~al.}(2001)\citenamefont {Bonn},
  \citenamefont {Wynn}, \citenamefont {Gardner}, \citenamefont {Lin},
  \citenamefont {Liang}, \citenamefont {Hardy}, \citenamefont {Kirtley},\ and\
  \citenamefont {Moler}}]{Bonn2001}%
  \BibitemOpen
  \bibfield  {author} {\bibinfo {author} {\bibfnamefont {D.~A.}\ \bibnamefont
  {Bonn}}, \bibinfo {author} {\bibfnamefont {Janice~C.}\ \bibnamefont {Wynn}},
  \bibinfo {author} {\bibfnamefont {Brian~W.}\ \bibnamefont {Gardner}},
  \bibinfo {author} {\bibfnamefont {Yu-Ju}\ \bibnamefont {Lin}}, \bibinfo
  {author} {\bibfnamefont {Ruixing}\ \bibnamefont {Liang}}, \bibinfo {author}
  {\bibfnamefont {W.~N.}\ \bibnamefont {Hardy}}, \bibinfo {author}
  {\bibfnamefont {J.~R.}\ \bibnamefont {Kirtley}}, \ and\ \bibinfo {author}
  {\bibfnamefont {K.~A.}\ \bibnamefont {Moler}},\ }\bibfield  {title} {\enquote
  {\bibinfo {title} {{A limit on spin–charge separation in high-T$_c$
  superconductors from the absence of a vortex-memory effect}},}\ }\href
  {\doibase 10.1038/414887a} {\bibfield  {journal} {\bibinfo  {journal}
  {Nature}\ }\textbf {\bibinfo {volume} {414}},\ \bibinfo {pages} {887--889}
  (\bibinfo {year} {2001})}\BibitemShut {NoStop}%
\bibitem [{\citenamefont {Persky}\ \emph {et~al.}(2022)\citenamefont {Persky},
  \citenamefont {Bjørlig}, \citenamefont {Feldman}, \citenamefont {Almoalem},
  \citenamefont {Altman}, \citenamefont {Berg}, \citenamefont {Kimchi},
  \citenamefont {Ruhman}, \citenamefont {Kanigel},\ and\ \citenamefont
  {Kalisky}}]{Persky2022}%
  \BibitemOpen
  \bibfield  {author} {\bibinfo {author} {\bibfnamefont {Eylon}\ \bibnamefont
  {Persky}}, \bibinfo {author} {\bibfnamefont {Anders~V.}\ \bibnamefont
  {Bjørlig}}, \bibinfo {author} {\bibfnamefont {Irena}\ \bibnamefont
  {Feldman}}, \bibinfo {author} {\bibfnamefont {Avior}\ \bibnamefont
  {Almoalem}}, \bibinfo {author} {\bibfnamefont {Ehud}\ \bibnamefont {Altman}},
  \bibinfo {author} {\bibfnamefont {Erez}\ \bibnamefont {Berg}}, \bibinfo
  {author} {\bibfnamefont {Itamar}\ \bibnamefont {Kimchi}}, \bibinfo {author}
  {\bibfnamefont {Jonathan}\ \bibnamefont {Ruhman}}, \bibinfo {author}
  {\bibfnamefont {Amit}\ \bibnamefont {Kanigel}}, \ and\ \bibinfo {author}
  {\bibfnamefont {Beena}\ \bibnamefont {Kalisky}},\ }\bibfield  {title}
  {\enquote {\bibinfo {title} {{Magnetic memory and spontaneous vortices in a
  van der Waals superconductor}},}\ }\href {\doibase
  10.1038/s41586-022-04855-2} {\bibfield  {journal} {\bibinfo  {journal}
  {Nature}\ }\textbf {\bibinfo {volume} {607}},\ \bibinfo {pages} {692--696}
  (\bibinfo {year} {2022})}\BibitemShut {NoStop}%
\bibitem [{\citenamefont {Shen}\ \emph
  {et~al.}(2022{\natexlab{a}})\citenamefont {Shen}, \citenamefont {Wen},
  \citenamefont {Kong}, \citenamefont {Gao}, \citenamefont {Si}, \citenamefont
  {Luo}, \citenamefont {Lu}, \citenamefont {Sun}, \citenamefont {Chen},\ and\
  \citenamefont {Yan}}]{Shen_2022}%
  \BibitemOpen
  \bibfield  {author} {\bibinfo {author} {\bibfnamefont {Shiwei}\ \bibnamefont
  {Shen}}, \bibinfo {author} {\bibfnamefont {Chenhaoping}\ \bibnamefont {Wen}},
  \bibinfo {author} {\bibfnamefont {Pengfei}\ \bibnamefont {Kong}}, \bibinfo
  {author} {\bibfnamefont {Jingjing}\ \bibnamefont {Gao}}, \bibinfo {author}
  {\bibfnamefont {Jianguo}\ \bibnamefont {Si}}, \bibinfo {author}
  {\bibfnamefont {Xuan}\ \bibnamefont {Luo}}, \bibinfo {author} {\bibfnamefont
  {Wenjian}\ \bibnamefont {Lu}}, \bibinfo {author} {\bibfnamefont {Yuping}\
  \bibnamefont {Sun}}, \bibinfo {author} {\bibfnamefont {Gang}\ \bibnamefont
  {Chen}}, \ and\ \bibinfo {author} {\bibfnamefont {Shichao}\ \bibnamefont
  {Yan}},\ }\bibfield  {title} {\enquote {\bibinfo {title} {Inducing and tuning
  kondo screening in a narrow-electronic-band system},}\ }\href {\doibase
  10.1038/s41467-022-29891-4} {\bibfield  {journal} {\bibinfo  {journal}
  {Nature Communications}\ }\textbf {\bibinfo {volume} {13}} (\bibinfo {year}
  {2022}{\natexlab{a}}),\ 10.1038/s41467-022-29891-4}\BibitemShut {NoStop}%
\bibitem [{\citenamefont {Hossain}\ \emph {et~al.}(2017)\citenamefont
  {Hossain}, \citenamefont {Zhao}, \citenamefont {Wen}, \citenamefont {Wang},
  \citenamefont {Wu},\ and\ \citenamefont {Xie}}]{cryst7100298}%
  \BibitemOpen
  \bibfield  {author} {\bibinfo {author} {\bibfnamefont {Mongur}\ \bibnamefont
  {Hossain}}, \bibinfo {author} {\bibfnamefont {Zhaoyang}\ \bibnamefont
  {Zhao}}, \bibinfo {author} {\bibfnamefont {Wen}\ \bibnamefont {Wen}},
  \bibinfo {author} {\bibfnamefont {Xinsheng}\ \bibnamefont {Wang}}, \bibinfo
  {author} {\bibfnamefont {Juanxia}\ \bibnamefont {Wu}}, \ and\ \bibinfo
  {author} {\bibfnamefont {Liming}\ \bibnamefont {Xie}},\ }\bibfield  {title}
  {\enquote {\bibinfo {title} {{Recent Advances in Two-Dimensional Materials
  with Charge Density Waves: Synthesis, Characterization and Applications}},}\
  }\href {\doibase 10.3390/cryst7100298} {\bibfield  {journal} {\bibinfo
  {journal} {Crystals}\ }\textbf {\bibinfo {volume} {7}} (\bibinfo {year}
  {2017}),\ 10.3390/cryst7100298}\BibitemShut {NoStop}%
\bibitem [{\citenamefont {Nayak}\ \emph {et~al.}(2021)\citenamefont {Nayak},
  \citenamefont {Steinbok}, \citenamefont {Roet}, \citenamefont {Koo},
  \citenamefont {Margalit}, \citenamefont {Feldman}, \citenamefont {Almoalem},
  \citenamefont {Kanigel}, \citenamefont {Fiete}, \citenamefont {Yan},
  \citenamefont {Oreg}, \citenamefont {Avraham},\ and\ \citenamefont
  {Beidenkopf}}]{Nayak_2021}%
  \BibitemOpen
  \bibfield  {author} {\bibinfo {author} {\bibfnamefont {Abhay~Kumar}\
  \bibnamefont {Nayak}}, \bibinfo {author} {\bibfnamefont {Aviram}\
  \bibnamefont {Steinbok}}, \bibinfo {author} {\bibfnamefont {Yotam}\
  \bibnamefont {Roet}}, \bibinfo {author} {\bibfnamefont {Jahyun}\ \bibnamefont
  {Koo}}, \bibinfo {author} {\bibfnamefont {Gilad}\ \bibnamefont {Margalit}},
  \bibinfo {author} {\bibfnamefont {Irena}\ \bibnamefont {Feldman}}, \bibinfo
  {author} {\bibfnamefont {Avior}\ \bibnamefont {Almoalem}}, \bibinfo {author}
  {\bibfnamefont {Amit}\ \bibnamefont {Kanigel}}, \bibinfo {author}
  {\bibfnamefont {Gregory~A.}\ \bibnamefont {Fiete}}, \bibinfo {author}
  {\bibfnamefont {Binghai}\ \bibnamefont {Yan}}, \bibinfo {author}
  {\bibfnamefont {Yuval}\ \bibnamefont {Oreg}}, \bibinfo {author}
  {\bibfnamefont {Nurit}\ \bibnamefont {Avraham}}, \ and\ \bibinfo {author}
  {\bibfnamefont {Haim}\ \bibnamefont {Beidenkopf}},\ }\bibfield  {title}
  {\enquote {\bibinfo {title} {Evidence of topological boundary modes with
  topological nodal-point superconductivity},}\ }\href {\doibase
  10.1038/s41567-021-01376-z} {\bibfield  {journal} {\bibinfo  {journal}
  {Nature Physics}\ }\textbf {\bibinfo {volume} {17}},\ \bibinfo {pages}
  {1413--1419} (\bibinfo {year} {2021})}\BibitemShut {NoStop}%
\bibitem [{\citenamefont {Law}\ and\ \citenamefont {Lee}(2017)}]{Law_2017}%
  \BibitemOpen
  \bibfield  {author} {\bibinfo {author} {\bibfnamefont {K.~T.}\ \bibnamefont
  {Law}}\ and\ \bibinfo {author} {\bibfnamefont {Patrick~A.}\ \bibnamefont
  {Lee}},\ }\bibfield  {title} {\enquote {\bibinfo {title} {{1T-{TaS}$_2$ as a
  quantum spin liquid}},}\ }\href {\doibase 10.1073/pnas.1706769114} {\bibfield
   {journal} {\bibinfo  {journal} {Proceedings of the National Academy of
  Sciences}\ }\textbf {\bibinfo {volume} {114}},\ \bibinfo {pages} {6996--7000}
  (\bibinfo {year} {2017})}\BibitemShut {NoStop}%
\bibitem [{\citenamefont {Li}\ \emph {et~al.}(2022)\citenamefont {Li},
  \citenamefont {Yao}, \citenamefont {Liu},\ and\ \citenamefont
  {Chen}}]{PhysRevLett.129.017202}%
  \BibitemOpen
  \bibfield  {author} {\bibinfo {author} {\bibfnamefont {Chao-Kai}\
  \bibnamefont {Li}}, \bibinfo {author} {\bibfnamefont {Xu-Ping}\ \bibnamefont
  {Yao}}, \bibinfo {author} {\bibfnamefont {Jianpeng}\ \bibnamefont {Liu}}, \
  and\ \bibinfo {author} {\bibfnamefont {Gang}\ \bibnamefont {Chen}},\
  }\bibfield  {title} {\enquote {\bibinfo {title} {{Fractionalization on the
  Surface: Is Type-II Terminated 1T-TaS$_2$ Surface an Anomalously Realized
  Spin Liquid?}}}\ }\href {\doibase 10.1103/PhysRevLett.129.017202} {\bibfield
  {journal} {\bibinfo  {journal} {Phys. Rev. Lett.}\ }\textbf {\bibinfo
  {volume} {129}},\ \bibinfo {pages} {017202} (\bibinfo {year}
  {2022})}\BibitemShut {NoStop}%
\bibitem [{\citenamefont {He}\ \emph {et~al.}(2018)\citenamefont {He},
  \citenamefont {Xu}, \citenamefont {Chen}, \citenamefont {Law},\ and\
  \citenamefont {Lee}}]{PhysRevLett.121.046401}%
  \BibitemOpen
  \bibfield  {author} {\bibinfo {author} {\bibfnamefont {Wen-Yu}\ \bibnamefont
  {He}}, \bibinfo {author} {\bibfnamefont {Xiao~Yan}\ \bibnamefont {Xu}},
  \bibinfo {author} {\bibfnamefont {Gang}\ \bibnamefont {Chen}}, \bibinfo
  {author} {\bibfnamefont {K.~T.}\ \bibnamefont {Law}}, \ and\ \bibinfo
  {author} {\bibfnamefont {Patrick~A.}\ \bibnamefont {Lee}},\ }\bibfield
  {title} {\enquote {\bibinfo {title} {{Spinon Fermi Surface in a Cluster Mott
  Insulator Model on a Triangular Lattice and Possible Application to
  1T-TaS$_2$}},}\ }\href {\doibase 10.1103/PhysRevLett.121.046401} {\bibfield
  {journal} {\bibinfo  {journal} {Phys. Rev. Lett.}\ }\textbf {\bibinfo
  {volume} {121}},\ \bibinfo {pages} {046401} (\bibinfo {year}
  {2018})}\BibitemShut {NoStop}%
\bibitem [{\citenamefont {Chen}\ and\ \citenamefont
  {Lee}(2018)}]{PhysRevB.97.035124}%
  \BibitemOpen
  \bibfield  {author} {\bibinfo {author} {\bibfnamefont {Gang}\ \bibnamefont
  {Chen}}\ and\ \bibinfo {author} {\bibfnamefont {Patrick~A.}\ \bibnamefont
  {Lee}},\ }\bibfield  {title} {\enquote {\bibinfo {title} {{Emergent orbitals
  in the cluster Mott insulator on a breathing kagome lattice}},}\ }\href
  {\doibase 10.1103/PhysRevB.97.035124} {\bibfield  {journal} {\bibinfo
  {journal} {Phys. Rev. B}\ }\textbf {\bibinfo {volume} {97}},\ \bibinfo
  {pages} {035124} (\bibinfo {year} {2018})}\BibitemShut {NoStop}%
\bibitem [{\citenamefont {Chen}\ \emph {et~al.}(2016)\citenamefont {Chen},
  \citenamefont {Kee},\ and\ \citenamefont {Kim}}]{PhysRevB.93.245134}%
  \BibitemOpen
  \bibfield  {author} {\bibinfo {author} {\bibfnamefont {Gang}\ \bibnamefont
  {Chen}}, \bibinfo {author} {\bibfnamefont {Hae-Young}\ \bibnamefont {Kee}}, \
  and\ \bibinfo {author} {\bibfnamefont {Yong~Baek}\ \bibnamefont {Kim}},\
  }\bibfield  {title} {\enquote {\bibinfo {title} {{Cluster Mott insulators and
  two Curie-Weiss regimes on an anisotropic kagome lattice}},}\ }\href
  {\doibase 10.1103/PhysRevB.93.245134} {\bibfield  {journal} {\bibinfo
  {journal} {Phys. Rev. B}\ }\textbf {\bibinfo {volume} {93}},\ \bibinfo
  {pages} {245134} (\bibinfo {year} {2016})}\BibitemShut {NoStop}%
\bibitem [{\citenamefont {Yao}\ \emph {et~al.}(2020)\citenamefont {Yao},
  \citenamefont {Zhang}, \citenamefont {Kim}, \citenamefont {Wang},\ and\
  \citenamefont {Chen}}]{PhysRevResearch.2.043424}%
  \BibitemOpen
  \bibfield  {author} {\bibinfo {author} {\bibfnamefont {Xu-Ping}\ \bibnamefont
  {Yao}}, \bibinfo {author} {\bibfnamefont {Xiao-Tian}\ \bibnamefont {Zhang}},
  \bibinfo {author} {\bibfnamefont {Yong~Baek}\ \bibnamefont {Kim}}, \bibinfo
  {author} {\bibfnamefont {Xiaoqun}\ \bibnamefont {Wang}}, \ and\ \bibinfo
  {author} {\bibfnamefont {Gang}\ \bibnamefont {Chen}},\ }\bibfield  {title}
  {\enquote {\bibinfo {title} {{Clusterization transition between cluster Mott
  insulators on a breathing kagome lattice}},}\ }\href {\doibase
  10.1103/PhysRevResearch.2.043424} {\bibfield  {journal} {\bibinfo  {journal}
  {Phys. Rev. Research}\ }\textbf {\bibinfo {volume} {2}},\ \bibinfo {pages}
  {043424} (\bibinfo {year} {2020})}\BibitemShut {NoStop}%
\bibitem [{\citenamefont {Shen}\ \emph
  {et~al.}(2022{\natexlab{b}})\citenamefont {Shen}, \citenamefont {Qin},
  \citenamefont {Gao}, \citenamefont {Wen}, \citenamefont {Wang}, \citenamefont
  {Wang}, \citenamefont {Li}, \citenamefont {Luo}, \citenamefont {Lu},
  \citenamefont {Sun},\ and\ \citenamefont {Yan}}]{Shichao2022}%
  \BibitemOpen
  \bibfield  {author} {\bibinfo {author} {\bibfnamefont {Shiwei}\ \bibnamefont
  {Shen}}, \bibinfo {author} {\bibfnamefont {Tian}\ \bibnamefont {Qin}},
  \bibinfo {author} {\bibfnamefont {Jingjing}\ \bibnamefont {Gao}}, \bibinfo
  {author} {\bibfnamefont {Chenhaoping}\ \bibnamefont {Wen}}, \bibinfo {author}
  {\bibfnamefont {Jinghui}\ \bibnamefont {Wang}}, \bibinfo {author}
  {\bibfnamefont {Wei}\ \bibnamefont {Wang}}, \bibinfo {author} {\bibfnamefont
  {Jun}\ \bibnamefont {Li}}, \bibinfo {author} {\bibfnamefont {Xuan}\
  \bibnamefont {Luo}}, \bibinfo {author} {\bibfnamefont {Wenjian}\ \bibnamefont
  {Lu}}, \bibinfo {author} {\bibfnamefont {Yuping}\ \bibnamefont {Sun}}, \ and\
  \bibinfo {author} {\bibfnamefont {Shichao}\ \bibnamefont {Yan}},\ }\bibfield
  {title} {\enquote {\bibinfo {title} {{Coexistence of Quasi-two-dimensional
  Superconductivity and Tunable Kondo Lattice in a van der Waals
  Superconductor}},}\ }\href {\doibase 10.1088/0256-307x/39/7/077401}
  {\bibfield  {journal} {\bibinfo  {journal} {Chinese Physics Letters}\
  }\textbf {\bibinfo {volume} {39}},\ \bibinfo {pages} {077401} (\bibinfo
  {year} {2022}{\natexlab{b}})}\BibitemShut {NoStop}%
\bibitem [{\citenamefont {Kurosaki}\ \emph {et~al.}(2005)\citenamefont
  {Kurosaki}, \citenamefont {Shimizu}, \citenamefont {Miyagawa}, \citenamefont
  {Kanoda},\ and\ \citenamefont {Saito}}]{PhysRevLett.95.177001}%
  \BibitemOpen
  \bibfield  {author} {\bibinfo {author} {\bibfnamefont {Y.}~\bibnamefont
  {Kurosaki}}, \bibinfo {author} {\bibfnamefont {Y.}~\bibnamefont {Shimizu}},
  \bibinfo {author} {\bibfnamefont {K.}~\bibnamefont {Miyagawa}}, \bibinfo
  {author} {\bibfnamefont {K.}~\bibnamefont {Kanoda}}, \ and\ \bibinfo {author}
  {\bibfnamefont {G.}~\bibnamefont {Saito}},\ }\bibfield  {title} {\enquote
  {\bibinfo {title} {{Mott Transition from a Spin Liquid to a Fermi Liquid in
  the Spin-Frustrated Organic Conductor $\kappa$-(ET)$_2$Cu$_2$(CN)$_3$ }},}\
  }\href {\doibase 10.1103/PhysRevLett.95.177001} {\bibfield  {journal}
  {\bibinfo  {journal} {Phys. Rev. Lett.}\ }\textbf {\bibinfo {volume} {95}},\
  \bibinfo {pages} {177001} (\bibinfo {year} {2005})}\BibitemShut {NoStop}%
\bibitem [{\citenamefont {Itou}\ \emph {et~al.}(2008)\citenamefont {Itou},
  \citenamefont {Oyamada}, \citenamefont {Maegawa}, \citenamefont {Tamura},\
  and\ \citenamefont {Kato}}]{PhysRevB.77.104413}%
  \BibitemOpen
  \bibfield  {author} {\bibinfo {author} {\bibfnamefont {T.}~\bibnamefont
  {Itou}}, \bibinfo {author} {\bibfnamefont {A.}~\bibnamefont {Oyamada}},
  \bibinfo {author} {\bibfnamefont {S.}~\bibnamefont {Maegawa}}, \bibinfo
  {author} {\bibfnamefont {M.}~\bibnamefont {Tamura}}, \ and\ \bibinfo {author}
  {\bibfnamefont {R.}~\bibnamefont {Kato}},\ }\bibfield  {title} {\enquote
  {\bibinfo {title} {{Quantum spin liquid in the spin-1/2 triangular
  antiferromagnet EtMe$_3$Sb[Pd(dmit)$_2$]$_2$ }},}\ }\href {\doibase
  10.1103/PhysRevB.77.104413} {\bibfield  {journal} {\bibinfo  {journal} {Phys.
  Rev. B}\ }\textbf {\bibinfo {volume} {77}},\ \bibinfo {pages} {104413}
  (\bibinfo {year} {2008})}\BibitemShut {NoStop}%
\bibitem [{\citenamefont {Klanj{\v{s}}ek}\ \emph {et~al.}(2017)\citenamefont
  {Klanj{\v{s}}ek}, \citenamefont {Zorko}, \citenamefont {{\v{Z}}itko},
  \citenamefont {Mravlje}, \citenamefont {Jagli{\v{c}}i{\'{c}}}, \citenamefont
  {Biswas}, \citenamefont {Prelov{\v{s}}ek}, \citenamefont {Mihailovic},\ and\
  \citenamefont {Ar{\v{c}}on}}]{Klanj_ek_2017}%
  \BibitemOpen
  \bibfield  {author} {\bibinfo {author} {\bibfnamefont {Martin}\ \bibnamefont
  {Klanj{\v{s}}ek}}, \bibinfo {author} {\bibfnamefont {Andrej}\ \bibnamefont
  {Zorko}}, \bibinfo {author} {\bibfnamefont {Rok}\ \bibnamefont
  {{\v{Z}}itko}}, \bibinfo {author} {\bibfnamefont {Jernej}\ \bibnamefont
  {Mravlje}}, \bibinfo {author} {\bibfnamefont {Zvonko}\ \bibnamefont
  {Jagli{\v{c}}i{\'{c}}}}, \bibinfo {author} {\bibfnamefont {Pabitra~Kumar}\
  \bibnamefont {Biswas}}, \bibinfo {author} {\bibfnamefont {Peter}\
  \bibnamefont {Prelov{\v{s}}ek}}, \bibinfo {author} {\bibfnamefont {Dragan}\
  \bibnamefont {Mihailovic}}, \ and\ \bibinfo {author} {\bibfnamefont {Denis}\
  \bibnamefont {Ar{\v{c}}on}},\ }\bibfield  {title} {\enquote {\bibinfo {title}
  {A high-temperature quantum spin liquid with polaron spins},}\ }\href
  {\doibase 10.1038/nphys4212} {\bibfield  {journal} {\bibinfo  {journal}
  {Nature Physics}\ }\textbf {\bibinfo {volume} {13}},\ \bibinfo {pages}
  {1130--1134} (\bibinfo {year} {2017})}\BibitemShut {NoStop}%
\bibitem [{\citenamefont {Ruan}\ \emph {et~al.}(2021)\citenamefont {Ruan},
  \citenamefont {Chen}, \citenamefont {Tang}, \citenamefont {Hwang},
  \citenamefont {Tsai}, \citenamefont {Lee}, \citenamefont {Wu}, \citenamefont
  {Ryu}, \citenamefont {Kahn}, \citenamefont {Liou}, \citenamefont {Jia},
  \citenamefont {Aikawa}, \citenamefont {Hwang}, \citenamefont {Wang},
  \citenamefont {Choi}, \citenamefont {Louie}, \citenamefont {Lee},
  \citenamefont {Shen}, \citenamefont {Mo},\ and\ \citenamefont
  {Crommie}}]{Ruan_2021}%
  \BibitemOpen
  \bibfield  {author} {\bibinfo {author} {\bibfnamefont {Wei}\ \bibnamefont
  {Ruan}}, \bibinfo {author} {\bibfnamefont {Yi}~\bibnamefont {Chen}}, \bibinfo
  {author} {\bibfnamefont {Shujie}\ \bibnamefont {Tang}}, \bibinfo {author}
  {\bibfnamefont {Jinwoong}\ \bibnamefont {Hwang}}, \bibinfo {author}
  {\bibfnamefont {Hsin-Zon}\ \bibnamefont {Tsai}}, \bibinfo {author}
  {\bibfnamefont {Ryan~L.}\ \bibnamefont {Lee}}, \bibinfo {author}
  {\bibfnamefont {Meng}\ \bibnamefont {Wu}}, \bibinfo {author} {\bibfnamefont
  {Hyejin}\ \bibnamefont {Ryu}}, \bibinfo {author} {\bibfnamefont {Salman}\
  \bibnamefont {Kahn}}, \bibinfo {author} {\bibfnamefont {Franklin}\
  \bibnamefont {Liou}}, \bibinfo {author} {\bibfnamefont {Caihong}\
  \bibnamefont {Jia}}, \bibinfo {author} {\bibfnamefont {Andrew}\ \bibnamefont
  {Aikawa}}, \bibinfo {author} {\bibfnamefont {Choongyu}\ \bibnamefont
  {Hwang}}, \bibinfo {author} {\bibfnamefont {Feng}\ \bibnamefont {Wang}},
  \bibinfo {author} {\bibfnamefont {Yongseong}\ \bibnamefont {Choi}}, \bibinfo
  {author} {\bibfnamefont {Steven~G.}\ \bibnamefont {Louie}}, \bibinfo {author}
  {\bibfnamefont {Patrick~A.}\ \bibnamefont {Lee}}, \bibinfo {author}
  {\bibfnamefont {Zhi-Xun}\ \bibnamefont {Shen}}, \bibinfo {author}
  {\bibfnamefont {Sung-Kwan}\ \bibnamefont {Mo}}, \ and\ \bibinfo {author}
  {\bibfnamefont {Michael~F.}\ \bibnamefont {Crommie}},\ }\bibfield  {title}
  {\enquote {\bibinfo {title} {{Evidence for quantum spin liquid behaviour in
  single-layer 1T-{TaSe}$_2$ from scanning tunnelling microscopy}},}\ }\href
  {\doibase 10.1038/s41567-021-01321-0} {\bibfield  {journal} {\bibinfo
  {journal} {Nature Physics}\ }\textbf {\bibinfo {volume} {17}},\ \bibinfo
  {pages} {1154--1161} (\bibinfo {year} {2021})}\BibitemShut {NoStop}%
\bibitem [{\citenamefont {Chen}\ \emph {et~al.}(2022)\citenamefont {Chen},
  \citenamefont {He}, \citenamefont {Ruan}, \citenamefont {Hwang},
  \citenamefont {Tang}, \citenamefont {Lee}, \citenamefont {Wu}, \citenamefont
  {Zhu}, \citenamefont {Zhang}, \citenamefont {Ryu}, \citenamefont {Wang},
  \citenamefont {Louie}, \citenamefont {Shen}, \citenamefont {Mo},
  \citenamefont {Lee},\ and\ \citenamefont {Crommie}}]{Chen_2022}%
  \BibitemOpen
  \bibfield  {author} {\bibinfo {author} {\bibfnamefont {Yi}~\bibnamefont
  {Chen}}, \bibinfo {author} {\bibfnamefont {Wen-Yu}\ \bibnamefont {He}},
  \bibinfo {author} {\bibfnamefont {Wei}\ \bibnamefont {Ruan}}, \bibinfo
  {author} {\bibfnamefont {Jinwoong}\ \bibnamefont {Hwang}}, \bibinfo {author}
  {\bibfnamefont {Shujie}\ \bibnamefont {Tang}}, \bibinfo {author}
  {\bibfnamefont {Ryan~L.}\ \bibnamefont {Lee}}, \bibinfo {author}
  {\bibfnamefont {Meng}\ \bibnamefont {Wu}}, \bibinfo {author} {\bibfnamefont
  {Tiancong}\ \bibnamefont {Zhu}}, \bibinfo {author} {\bibfnamefont {Canxun}\
  \bibnamefont {Zhang}}, \bibinfo {author} {\bibfnamefont {Hyejin}\
  \bibnamefont {Ryu}}, \bibinfo {author} {\bibfnamefont {Feng}\ \bibnamefont
  {Wang}}, \bibinfo {author} {\bibfnamefont {Steven~G.}\ \bibnamefont {Louie}},
  \bibinfo {author} {\bibfnamefont {Zhi-Xun}\ \bibnamefont {Shen}}, \bibinfo
  {author} {\bibfnamefont {Sung-Kwan}\ \bibnamefont {Mo}}, \bibinfo {author}
  {\bibfnamefont {Patrick~A.}\ \bibnamefont {Lee}}, \ and\ \bibinfo {author}
  {\bibfnamefont {Michael~F.}\ \bibnamefont {Crommie}},\ }\bibfield  {title}
  {\enquote {\bibinfo {title} {{Evidence for a spinon Kondo effect in cobalt
  atoms on single-layer 1T-TaSe$_2$}},}\ }\href {\doibase
  10.1038/s41567-022-01751-4} {\bibfield  {journal} {\bibinfo  {journal}
  {Nature Physics}\ }\textbf {\bibinfo {volume} {18}},\ \bibinfo {pages}
  {1335–1340} (\bibinfo {year} {2022})}\BibitemShut {NoStop}%
\bibitem [{\citenamefont {Qiao}\ \emph {et~al.}(2017)\citenamefont {Qiao},
  \citenamefont {Li}, \citenamefont {Wang}, \citenamefont {Ruan}, \citenamefont
  {Ye}, \citenamefont {Cai}, \citenamefont {Hao}, \citenamefont {Yao},
  \citenamefont {Chen}, \citenamefont {Wu}, \citenamefont {Wang},\ and\
  \citenamefont {Liu}}]{PhysRevX.7.041054}%
  \BibitemOpen
  \bibfield  {author} {\bibinfo {author} {\bibfnamefont {Shuang}\ \bibnamefont
  {Qiao}}, \bibinfo {author} {\bibfnamefont {Xintong}\ \bibnamefont {Li}},
  \bibinfo {author} {\bibfnamefont {Naizhou}\ \bibnamefont {Wang}}, \bibinfo
  {author} {\bibfnamefont {Wei}\ \bibnamefont {Ruan}}, \bibinfo {author}
  {\bibfnamefont {Cun}\ \bibnamefont {Ye}}, \bibinfo {author} {\bibfnamefont
  {Peng}\ \bibnamefont {Cai}}, \bibinfo {author} {\bibfnamefont {Zhenqi}\
  \bibnamefont {Hao}}, \bibinfo {author} {\bibfnamefont {Hong}\ \bibnamefont
  {Yao}}, \bibinfo {author} {\bibfnamefont {Xianhui}\ \bibnamefont {Chen}},
  \bibinfo {author} {\bibfnamefont {Jian}\ \bibnamefont {Wu}}, \bibinfo
  {author} {\bibfnamefont {Yayu}\ \bibnamefont {Wang}}, \ and\ \bibinfo
  {author} {\bibfnamefont {Zheng}\ \bibnamefont {Liu}},\ }\bibfield  {title}
  {\enquote {\bibinfo {title} {{Mottness Collapse in 1T-TaS$_{2-x}$Se$_{x}$
  Transition-Metal Dichalcogenide: An Interplay between Localized and Itinerant
  Orbitals}},}\ }\href {\doibase 10.1103/PhysRevX.7.041054} {\bibfield
  {journal} {\bibinfo  {journal} {Phys. Rev. X}\ }\textbf {\bibinfo {volume}
  {7}},\ \bibinfo {pages} {041054} (\bibinfo {year} {2017})}\BibitemShut
  {NoStop}%
\bibitem [{\citenamefont {{Ang}}\ \emph {et~al.}(2015)\citenamefont {{Ang}},
  \citenamefont {{Wang}}, \citenamefont {{Chen}}, \citenamefont {{Tang}},
  \citenamefont {{Liu}}, \citenamefont {{Liu}}, \citenamefont {{Lu}},
  \citenamefont {{Sun}}, \citenamefont {{Mori}},\ and\ \citenamefont
  {{Ikuhara}}}]{2015NatCo...6.6091A}%
  \BibitemOpen
  \bibfield  {author} {\bibinfo {author} {\bibfnamefont {R.}~\bibnamefont
  {{Ang}}}, \bibinfo {author} {\bibfnamefont {Z.~C.}\ \bibnamefont {{Wang}}},
  \bibinfo {author} {\bibfnamefont {C.~L.}\ \bibnamefont {{Chen}}}, \bibinfo
  {author} {\bibfnamefont {J.}~\bibnamefont {{Tang}}}, \bibinfo {author}
  {\bibfnamefont {N.}~\bibnamefont {{Liu}}}, \bibinfo {author} {\bibfnamefont
  {Y.}~\bibnamefont {{Liu}}}, \bibinfo {author} {\bibfnamefont {W.~J.}\
  \bibnamefont {{Lu}}}, \bibinfo {author} {\bibfnamefont {Y.~P.}\ \bibnamefont
  {{Sun}}}, \bibinfo {author} {\bibfnamefont {T.}~\bibnamefont {{Mori}}}, \
  and\ \bibinfo {author} {\bibfnamefont {Y.}~\bibnamefont {{Ikuhara}}},\
  }\bibfield  {title} {\enquote {\bibinfo {title} {{Atomistic origin of an
  ordered superstructure induced superconductivity in layered
  chalcogenides}},}\ }\href {\doibase 10.1038/ncomms7091} {\bibfield  {journal}
  {\bibinfo  {journal} {Nature Communications}\ }\textbf {\bibinfo {volume}
  {6}},\ \bibinfo {eid} {6091} (\bibinfo {year} {2015})}\BibitemShut {NoStop}%
\bibitem [{\citenamefont {Motrunich}(2005)}]{PhysRevB.72.045105}%
  \BibitemOpen
  \bibfield  {author} {\bibinfo {author} {\bibfnamefont {Olexei~I.}\
  \bibnamefont {Motrunich}},\ }\bibfield  {title} {\enquote {\bibinfo {title}
  {{Variational study of triangular lattice spin-$1/2$ model with ring
  exchanges and spin liquid state in $\kappa$-(ET)$_2$Cu$_2$(CN)$_3$}},}\
  }\href {\doibase 10.1103/PhysRevB.72.045105} {\bibfield  {journal} {\bibinfo
  {journal} {Phys. Rev. B}\ }\textbf {\bibinfo {volume} {72}},\ \bibinfo
  {pages} {045105} (\bibinfo {year} {2005})}\BibitemShut {NoStop}%
\bibitem [{\citenamefont {Mishmash}\ \emph {et~al.}(2013)\citenamefont
  {Mishmash}, \citenamefont {Garrison}, \citenamefont {Bieri},\ and\
  \citenamefont {Xu}}]{PhysRevLett.111.157203}%
  \BibitemOpen
  \bibfield  {author} {\bibinfo {author} {\bibfnamefont {Ryan~V.}\ \bibnamefont
  {Mishmash}}, \bibinfo {author} {\bibfnamefont {James~R.}\ \bibnamefont
  {Garrison}}, \bibinfo {author} {\bibfnamefont {Samuel}\ \bibnamefont
  {Bieri}}, \ and\ \bibinfo {author} {\bibfnamefont {Cenke}\ \bibnamefont
  {Xu}},\ }\bibfield  {title} {\enquote {\bibinfo {title} {{Theory of a
  Competitive Spin Liquid State for Weak Mott Insulators on the Triangular
  Lattice}},}\ }\href {\doibase 10.1103/PhysRevLett.111.157203} {\bibfield
  {journal} {\bibinfo  {journal} {Phys. Rev. Lett.}\ }\textbf {\bibinfo
  {volume} {111}},\ \bibinfo {pages} {157203} (\bibinfo {year}
  {2013})}\BibitemShut {NoStop}%
\bibitem [{\citenamefont {Florens}\ and\ \citenamefont
  {Georges}(2004)}]{PhysRevB.70.035114}%
  \BibitemOpen
  \bibfield  {author} {\bibinfo {author} {\bibfnamefont {Serge}\ \bibnamefont
  {Florens}}\ and\ \bibinfo {author} {\bibfnamefont {Antoine}\ \bibnamefont
  {Georges}},\ }\bibfield  {title} {\enquote {\bibinfo {title} {{Slave-rotor
  mean-field theories of strongly correlated systems and the Mott transition in
  finite dimensions}},}\ }\href {\doibase 10.1103/PhysRevB.70.035114}
  {\bibfield  {journal} {\bibinfo  {journal} {Phys. Rev. B}\ }\textbf {\bibinfo
  {volume} {70}},\ \bibinfo {pages} {035114} (\bibinfo {year}
  {2004})}\BibitemShut {NoStop}%
\bibitem [{\citenamefont {Lee}\ and\ \citenamefont
  {Lee}(2005)}]{PhysRevLett.95.036403}%
  \BibitemOpen
  \bibfield  {author} {\bibinfo {author} {\bibfnamefont {Sung-Sik}\
  \bibnamefont {Lee}}\ and\ \bibinfo {author} {\bibfnamefont {Patrick~A.}\
  \bibnamefont {Lee}},\ }\bibfield  {title} {\enquote {\bibinfo {title} {{U(1)
  Gauge Theory of the Hubbard Model: Spin Liquid States and Possible
  Application to $\kappa$-(ET)$_2$Cu$_2$(CN)$_3$}},}\ }\href {\doibase
  10.1103/PhysRevLett.95.036403} {\bibfield  {journal} {\bibinfo  {journal}
  {Phys. Rev. Lett.}\ }\textbf {\bibinfo {volume} {95}},\ \bibinfo {pages}
  {036403} (\bibinfo {year} {2005})}\BibitemShut {NoStop}%
\bibitem [{\citenamefont {Szasz}\ \emph {et~al.}(2020)\citenamefont {Szasz},
  \citenamefont {Motruk}, \citenamefont {Zaletel},\ and\ \citenamefont
  {Moore}}]{PhysRevX.10.021042}%
  \BibitemOpen
  \bibfield  {author} {\bibinfo {author} {\bibfnamefont {Aaron}\ \bibnamefont
  {Szasz}}, \bibinfo {author} {\bibfnamefont {Johannes}\ \bibnamefont
  {Motruk}}, \bibinfo {author} {\bibfnamefont {Michael~P.}\ \bibnamefont
  {Zaletel}}, \ and\ \bibinfo {author} {\bibfnamefont {Joel~E.}\ \bibnamefont
  {Moore}},\ }\bibfield  {title} {\enquote {\bibinfo {title} {{Chiral Spin
  Liquid Phase of the Triangular Lattice Hubbard Model: A Density Matrix
  Renormalization Group Study}},}\ }\href {\doibase 10.1103/PhysRevX.10.021042}
  {\bibfield  {journal} {\bibinfo  {journal} {Phys. Rev. X}\ }\textbf {\bibinfo
  {volume} {10}},\ \bibinfo {pages} {021042} (\bibinfo {year}
  {2020})}\BibitemShut {NoStop}%
\bibitem [{\citenamefont {Cookmeyer}\ \emph {et~al.}(2021)\citenamefont
  {Cookmeyer}, \citenamefont {Motruk},\ and\ \citenamefont
  {Moore}}]{PhysRevLett.127.087201}%
  \BibitemOpen
  \bibfield  {author} {\bibinfo {author} {\bibfnamefont {Tessa}\ \bibnamefont
  {Cookmeyer}}, \bibinfo {author} {\bibfnamefont {Johannes}\ \bibnamefont
  {Motruk}}, \ and\ \bibinfo {author} {\bibfnamefont {Joel~E.}\ \bibnamefont
  {Moore}},\ }\bibfield  {title} {\enquote {\bibinfo {title} {{Four-Spin Terms
  and the Origin of the Chiral Spin Liquid in Mott Insulators on the Triangular
  Lattice}},}\ }\href {\doibase 10.1103/PhysRevLett.127.087201} {\bibfield
  {journal} {\bibinfo  {journal} {Phys. Rev. Lett.}\ }\textbf {\bibinfo
  {volume} {127}},\ \bibinfo {pages} {087201} (\bibinfo {year}
  {2021})}\BibitemShut {NoStop}%
\bibitem [{\citenamefont {Chen}(2017{\natexlab{a}})}]{PhysRevB.96.085136}%
  \BibitemOpen
  \bibfield  {author} {\bibinfo {author} {\bibfnamefont {Gang}\ \bibnamefont
  {Chen}},\ }\bibfield  {title} {\enquote {\bibinfo {title} {Spectral
  periodicity of the spinon continuum in quantum spin ice},}\ }\href {\doibase
  10.1103/PhysRevB.96.085136} {\bibfield  {journal} {\bibinfo  {journal} {Phys.
  Rev. B}\ }\textbf {\bibinfo {volume} {96}},\ \bibinfo {pages} {085136}
  (\bibinfo {year} {2017}{\natexlab{a}})}\BibitemShut {NoStop}%
\bibitem [{\citenamefont {Chen}(2017{\natexlab{b}})}]{PhysRevB.96.195127}%
  \BibitemOpen
  \bibfield  {author} {\bibinfo {author} {\bibfnamefont {Gang}\ \bibnamefont
  {Chen}},\ }\bibfield  {title} {\enquote {\bibinfo {title} {{Dirac's
  ``magnetic monopoles'' in pyrochlore ice $U(1)$ spin liquids: Spectrum and
  classification}},}\ }\href {\doibase 10.1103/PhysRevB.96.195127} {\bibfield
  {journal} {\bibinfo  {journal} {Phys. Rev. B}\ }\textbf {\bibinfo {volume}
  {96}},\ \bibinfo {pages} {195127} (\bibinfo {year}
  {2017}{\natexlab{b}})}\BibitemShut {NoStop}%
\bibitem [{\citenamefont {Essin}\ and\ \citenamefont
  {Hermele}(2014)}]{PhysRevB.90.121102}%
  \BibitemOpen
  \bibfield  {author} {\bibinfo {author} {\bibfnamefont {Andrew~M.}\
  \bibnamefont {Essin}}\ and\ \bibinfo {author} {\bibfnamefont {Michael}\
  \bibnamefont {Hermele}},\ }\bibfield  {title} {\enquote {\bibinfo {title}
  {Spectroscopic signatures of crystal momentum fractionalization},}\ }\href
  {\doibase 10.1103/PhysRevB.90.121102} {\bibfield  {journal} {\bibinfo
  {journal} {Phys. Rev. B}\ }\textbf {\bibinfo {volume} {90}},\ \bibinfo
  {pages} {121102} (\bibinfo {year} {2014})}\BibitemShut {NoStop}%
\bibitem [{\citenamefont {Guo}\ \emph {et~al.}(2020)\citenamefont {Guo},
  \citenamefont {Samajdar}, \citenamefont {Scheurer},\ and\ \citenamefont
  {Sachdev}}]{PhysRevB.101.195126}%
  \BibitemOpen
  \bibfield  {author} {\bibinfo {author} {\bibfnamefont {Haoyu}\ \bibnamefont
  {Guo}}, \bibinfo {author} {\bibfnamefont {Rhine}\ \bibnamefont {Samajdar}},
  \bibinfo {author} {\bibfnamefont {Mathias~S.}\ \bibnamefont {Scheurer}}, \
  and\ \bibinfo {author} {\bibfnamefont {Subir}\ \bibnamefont {Sachdev}},\
  }\bibfield  {title} {\enquote {\bibinfo {title} {{Gauge theories for the
  thermal Hall effect}},}\ }\href {\doibase 10.1103/PhysRevB.101.195126}
  {\bibfield  {journal} {\bibinfo  {journal} {Phys. Rev. B}\ }\textbf {\bibinfo
  {volume} {101}},\ \bibinfo {pages} {195126} (\bibinfo {year}
  {2020})}\BibitemShut {NoStop}%
\bibitem [{\citenamefont {Ribak}\ \emph {et~al.}(2020)\citenamefont {Ribak},
  \citenamefont {Skiff}, \citenamefont {Mograbi}, \citenamefont {Rout},
  \citenamefont {Fischer}, \citenamefont {Ruhman}, \citenamefont {Chashka},
  \citenamefont {Dagan},\ and\ \citenamefont {Kanigel}}]{sciadv.aax9480}%
  \BibitemOpen
  \bibfield  {author} {\bibinfo {author} {\bibfnamefont {A.}~\bibnamefont
  {Ribak}}, \bibinfo {author} {\bibfnamefont {R.~Majlin}\ \bibnamefont
  {Skiff}}, \bibinfo {author} {\bibfnamefont {M.}~\bibnamefont {Mograbi}},
  \bibinfo {author} {\bibfnamefont {P.~K.}\ \bibnamefont {Rout}}, \bibinfo
  {author} {\bibfnamefont {M.~H.}\ \bibnamefont {Fischer}}, \bibinfo {author}
  {\bibfnamefont {J.}~\bibnamefont {Ruhman}}, \bibinfo {author} {\bibfnamefont
  {K.}~\bibnamefont {Chashka}}, \bibinfo {author} {\bibfnamefont
  {Y.}~\bibnamefont {Dagan}}, \ and\ \bibinfo {author} {\bibfnamefont
  {A.}~\bibnamefont {Kanigel}},\ }\bibfield  {title} {\enquote {\bibinfo
  {title} {{Chiral superconductivity in the alternate stacking compound
  4Hb-TaS$_2$}},}\ }\href {\doibase 10.1126/sciadv.aax9480} {\bibfield
  {journal} {\bibinfo  {journal} {Science Advances}\ }\textbf {\bibinfo
  {volume} {6}},\ \bibinfo {pages} {eaax9480} (\bibinfo {year} {2020})},\
  \Eprint
  {http://arxiv.org/abs/https://www.science.org/doi/pdf/10.1126/sciadv.aax9480}
  {https://www.science.org/doi/pdf/10.1126/sciadv.aax9480} \BibitemShut
  {NoStop}%
\end{thebibliography}%

% referee: Patrick Lee, Senthil, Fisher, Paramekanti, Motrunich

\end{document}